%% file: template.tex
\RequirePackage{fix-cm}
\documentclass[twocolumn]{svjour3}          
\smartqed  
\usepackage{graphicx}
\usepackage{lineno}  
\usepackage{graphicx}  
\usepackage{xspace} 
\usepackage{color}
\usepackage{colortbl}
\usepackage{amsmath} 
\allowdisplaybreaks
\usepackage{ifthen} 
\newboolean{articletitles}
\setboolean{articletitles}{true} 

\newboolean{uprightparticles}
\setboolean{uprightparticles}{false} 
\input{lhcb-symbols-def}

\input{additional-symbols}
\begin{document}

\title{Experimental constraints from flavour changing processes and physics beyond the Standard Model\thanks{MG and VVG are supported by a Marie Curie Action: ``Cofunding of the CERN Fellowship Programme (COFUND-CERN)'' of the European Community's Seventh Framework Programme under contract number (PCOFUND-GA-2008-229600).}
}


\author{M. Gersabeck         \and
        V.V. Gligorov \and 
        N. Serra 
}


\institute{M. Gersabeck \at
              CERN, 1211 Geneva, Switzerland\\
              \email{marco.gersabeck@cern.ch}           
           \and
           V.V. Gligorov\at
             CERN, 1211 Geneva, Switzerland\\
	     \email{vladimir.gligorov@cern.ch}
	   \and
	   N. Serra\at
	     University of Zuerich, 8006 Zuerich, Switzerland\\
	     \email{nicola.serra@cern.ch}
}

\date{Received: date / Accepted: date}

\maketitle

\begin{abstract}
Flavour physics has a long tradition of paving the way for direct discoveries of new particles and interactions.
Results over the last decade have placed stringent bounds on the parameter space of physics beyond the Standard Model.
Early results from the \lhc, and its dedicated flavour factory \lhcb,
have further tightened these constraints and reiterate
the ongoing relevance of flavour studies. The experimental status of flavour observables in the charm and beauty sectors is reviewed
in measurements of \CP violation, neutral meson mixing, and measurements of rare decays.
\end{abstract}

\section{Introduction}
Flavour physics has given key contributions to the understanding of fundamental particles.
The kaon system is an excellent example how the interplay of meson anti-meson mixing~\cite{Lande:1956pf,Jackson:1957zzb,Niebergall:1974wh},
and the search for rare decays~\cite{BottBodenhausen1967194,Foeth:1969hi} led to the prediction of the charm quark and
indeed charm mesons~\cite{GellMann1964214,Tarjanne:1963zz,Hara:1963gw,Bjorken:1964gz,Glashow:1970gm}.
Furthermore, the observation of \CP violation in neutral kaons~\cite{Christenson:1964fg} led to the prediction of a
third generation of quarks~\cite{Kobayashi:1973fv}.
At the LHC,  precision measurements of flavour physics are sensitive to 
new particles contributing to quantum loops up to scales 
of about $200\tev$~\cite{Buras:2009if} which,
according to the Heisenberg uncertainty principle~\cite{Heisenberg:1927zz}, correspond to distance scales of the order of $10^{-21}~{\rm m}$.
This exceeds the reach for direct production of particles by roughly two orders of magnitude.

This review covers flavour changing processes of charm and beauty mesons; recent results on lepton flavour violating decays are also briefly discussed.
These provide complementary access to effects from Physics Beyond the Standard Model (PBSM).
This complementarity will eventually help to identify the nature of
signs of 
new dynamics, should they be generated by a common source.
Sections~\ref{s:cpv_intro} to~\ref{sec:cpv_beauty} cover the status of mixing and \CP violation
measurements while section~\ref{sec:rare} reviews measurements of rare decays. 
\input{cpv_intro}
\input{charm}
\input{BCPV}
\input{RareDecays}
\section{Conclusion}
Despite the ongoing lack of a direct discovery of particles beyond the Standard Model, recent results in flavour physics
are giving ever stronger hints of effects beyond the Standard Model. In particular, the observation of permille-level \CP violation
in \Dz decays, the large dimuon asymmetry in \Bz and \Bs decays, as well as the values of $\sin\left(2\beta\right)$ and the branching ratio of
$\Bp\to \taup \nu$, are difficult to simultaneously interpret within the Standard Model framework.
At the same time, measurements of rare decays such as $\Bs\to\mup\mu^-$ and $\Bz\to\Kstarz\mup\mu^-$ which are in good agreement with
the Standard Model have placed the most stringent limits yet on many Standard Model extensions. What this contradiction highlights
is the ongoing relevance of flavour physics as key tool not only for
the indirect discovery of new particles and processes, but also for
discriminating between the many proposed theories of physics 
beyond the Standard Model. With the excellent performance of the \lhc,
and the wealth of precision flavour measurements coming from its detectors, it is reasonable to hope for a deepening, and eventual resolution,
of these contradictions in the years to come.
\begin{acknowledgements}
The authors would like to thank the CKMFitter, HFAG, and UTFit collaborations for preparing 
updated world averages of many flavour parameters in time for this review, and Mitesh Patel for useful
comments on the draft. 
\end{acknowledgements}

\bibliographystyle{spphys}       
\bibliography{template}   

\end{document}

%% file: lhcb-symbols-def.tex
\def\lhcb {LHCb\xspace}
\def\ux85 {UX85\xspace}

\def\lhc {LHC\xspace}
\def\atlas {ATLAS\xspace}
\def\cms {CMS\xspace}
\def\babar  {BaBar\xspace}
\def\belle  {Belle\xspace}

\def\cdf    {CDF\xspace}
\def\dzero  {D\O\xspace}

\def\cleo   {CLEO\xspace}

\ifthenelse{\boolean{uprightparticles}}%
{

 \def\Pmu         {\ensuremath{\upmu}\xspace}                 
 \def\Pnu         {\ensuremath{\upnu}\xspace}                 
                  
 \def\Ppi         {\ensuremath{\uppi}\xspace}

 \def\Ptau        {\ensuremath{\uptau}\xspace}                 
                  
 \def\Pphi        {\ensuremath{\upphi}\xspace}

 \def\Ppsi        {\ensuremath{\uppsi}\xspace}

 \def\PDelta      {\ensuremath{\Delta}\xspace}                 
 \def\PXi      {\ensuremath{\Xi}\xspace}                 
 \def\PLambda      {\ensuremath{\Lambda}\xspace}                 
 \def\PSigma      {\ensuremath{\Sigma}\xspace}                 
 \def\POmega      {\ensuremath{\Omega}\xspace}                 
 \def\PUpsilon      {\ensuremath{\Upsilon}\xspace}                 
 
 \def\PB      {\ensuremath{\mathrm{B}}\xspace}                 
                  
 \def\PD      {\ensuremath{\mathrm{D}}\xspace}

 \def\PJ      {\ensuremath{\mathrm{J}}\xspace}                 
 \def\PK      {\ensuremath{\mathrm{K}}\xspace}                 
                  
 \def\PM      {\ensuremath{\mathrm{M}}\xspace}

 \def\Pb      {\ensuremath{\mathrm{b}}\xspace}                 
 \def\Pc      {\ensuremath{\mathrm{c}}\xspace}                 
                  
 \def\Pe      {\ensuremath{\mathrm{e}}\xspace}

 \def\Pi      {\ensuremath{\mathrm{i}}\xspace}

 \def\Ps      {\ensuremath{\mathrm{s}}\xspace}                 
                  
 \def\Pu      {\ensuremath{\mathrm{u}}\xspace}

}
{

 \def\Pmu         {\ensuremath{\mu}\xspace}                 
 \def\Pnu         {\ensuremath{\nu}\xspace}                 
                  
 \def\Ppi         {\ensuremath{\pi}\xspace}

 \def\Ptau        {\ensuremath{\tau}\xspace}                 
                  
 \def\Pphi        {\ensuremath{\phi}\xspace}

 \def\Ppsi        {\ensuremath{\psi}\xspace}                 
                  
 \mathchardef\PDelta="7101
 \mathchardef\PXi="7104
 \mathchardef\PLambda="7103
 \mathchardef\PSigma="7106
 \mathchardef\POmega="710A
 \mathchardef\PUpsilon="7107
                  
 \def\PB      {\ensuremath{B}\xspace}                 
                  
 \def\PD      {\ensuremath{D}\xspace}

 \def\PJ      {\ensuremath{J}\xspace}                 
 \def\PK      {\ensuremath{K}\xspace}                 
                  
 \def\PM      {\ensuremath{M}\xspace}

 \def\Pb      {\ensuremath{b}\xspace}                 
 \def\Pc      {\ensuremath{c}\xspace}                 
                  
 \def\Pe      {\ensuremath{e}\xspace}

 \def\Pi      {\ensuremath{i}\xspace}

 \def\Ps      {\ensuremath{s}\xspace}                 
                  
 \def\Pu      {\ensuremath{u}\xspace}

}

\def\en         {\ensuremath{\Pe^-}\xspace}   
\def\ep         {\ensuremath{\Pe^+}\xspace}

\def\mup        {\ensuremath{\Pmu^+}\xspace}
\def\mun        {\ensuremath{\Pmu^-}\xspace} 

\def\taup       {\ensuremath{\Ptau^+}\xspace}

\def\neu        {\ensuremath{\Pnu}\xspace}
\def\neub       {\ensuremath{\overline{\Pnu}}\xspace}

\def\neum       {\ensuremath{\neu_\mu}\xspace}
\def\neumb      {\ensuremath{\neub_\mu}\xspace}

\def\uquark    {\ensuremath{\Pu}\xspace}

\def\squark    {\ensuremath{\Ps}\xspace}

\def\cquark    {\ensuremath{\Pc}\xspace}

\def\bquark    {\ensuremath{\Pb}\xspace}

\def\pion  {\ensuremath{\Ppi}\xspace}
\def\piz   {\ensuremath{\pion^0}\xspace}

\def\pip   {\ensuremath{\pion^+}\xspace}
\def\pim   {\ensuremath{\pion^-}\xspace}

\def\kaon  {\ensuremath{\PK}\xspace}
  \def\Kbar  {\kern 0.2em\overline{\kern -0.2em \PK}{}\xspace}

\def\Kz    {\ensuremath{\kaon^0}\xspace}
\def\Kzb   {\ensuremath{\Kbar^0}\xspace}
\def\KzKzb {\ensuremath{\Kz \kern -0.16em \Kzb}\xspace}
\def\Kp    {\ensuremath{\kaon^+}\xspace}
\def\Km    {\ensuremath{\kaon^-}\xspace}

\def\KpKm  {\ensuremath{\Kp \kern -0.16em \Km}\xspace}
\def\KS    {\ensuremath{\kaon^0_{\rm\scriptscriptstyle S}}\xspace} 
 
\def\Kstarz  {\ensuremath{\kaon^{*0}}\xspace}

  \def\Dbar    {\kern 0.2em\overline{\kern -0.2em \PD}{}\xspace}
\def\D       {\ensuremath{\PD}\xspace}

\def\Dz      {\ensuremath{\D^0}\xspace}
\def\Dzb     {\ensuremath{\Dbar^0}\xspace}
\def\DzDzb   {\ensuremath{\Dz {\kern -0.16em \Dzb}}\xspace}
\def\Dp      {\ensuremath{\D^+}\xspace}
\def\Dm      {\ensuremath{\D^-}\xspace}

\def\DpDm    {\ensuremath{\Dp {\kern -0.16em \Dm}}\xspace}

\def\Dstarp  {\ensuremath{\D^{*+}}\xspace}

\def\Ds      {\ensuremath{\D^+_\squark}\xspace}

\def\Dspm    {\ensuremath{\D^{\pm}_\squark}\xspace}

\def\B       {\ensuremath{\PB}\xspace}
  \def\Bbar    {\kern 0.18em\overline{\kern -0.18em \PB}{}\xspace}

\def\Bz      {\ensuremath{\B^0}\xspace}

\def\Bu      {\ensuremath{\B^+}\xspace}
\def\Bub     {\ensuremath{\B^-}\xspace}
\def\Bp      {\ensuremath{\Bu}\xspace}
\def\Bm      {\ensuremath{\Bub}\xspace}
\def\Bpm     {\ensuremath{\B^\pm}\xspace}

\def\Bd      {\ensuremath{\B^0}\xspace}
\def\Bs      {\ensuremath{\B^0_\squark}\xspace}

\def\jpsi     {\ensuremath{{\PJ\mskip -3mu/\mskip -2mu\Ppsi\mskip 2mu}}\xspace}

  \def\Y#1S{\ensuremath{\PUpsilon{(#1S)}}\xspace}

\def\Lbar {\ensuremath{\kern 0.1em\overline{\kern -0.1em\Lambda\kern -0.05em}\kern 0.05em{}}\xspace}

\newcommand{\decay}[2]{\ensuremath{#1\!\to #2}\xspace}         

\def\to                 {\ensuremath{\rightarrow}\xspace}

\def\CP                {\ensuremath{C\!P}\xspace}

\def\Vub  {\ensuremath{|V_{\uquark\bquark}|}\xspace}
\def\Vcb  {\ensuremath{|V_{\cquark\bquark}|}\xspace}

\def\BsToJPsiPhi  {\decay{\Bs}{\jpsi\phi}}

\def\AFB      {\ensuremath{A_{\mathrm{FB}}}\xspace}

\def\AT#1     {\ensuremath{A_{\mathrm{T}}^{#1}}\xspace}           

\def\C#1      {\ensuremath{\mathcal{C}_{#1}}\xspace}                       
\def\Cp#1     {\ensuremath{\mathcal{C}_{#1}^{'}}\xspace}                    
\def\Ceff#1   {\ensuremath{\mathcal{C}_{#1}^{\mathrm{(eff)}}}\xspace}        
\def\Cpeff#1  {\ensuremath{\mathcal{C}_{#1}^{'\mathrm{(eff)}}}\xspace}       
\def\Ope#1    {\ensuremath{\mathcal{O}_{#1}}\xspace}                       
\def\Opep#1   {\ensuremath{\mathcal{O}_{#1}^{'}}\xspace}                    

\def\ycp        {\ensuremath{y_{\CP}}\xspace}
\def\agamma     {\ensuremath{A_{\Gamma}}\xspace}

\newcommand{\tev}{\ensuremath{\mathrm{\,Te\kern -0.1em V}}\xspace}
\newcommand{\gev}{\ensuremath{\mathrm{\,Ge\kern -0.1em V}}\xspace}
\newcommand{\mev}{\ensuremath{\mathrm{\,Me\kern -0.1em V}}\xspace}
\newcommand{\kev}{\ensuremath{\mathrm{\,ke\kern -0.1em V}}\xspace}
\newcommand{\ev}{\ensuremath{\mathrm{\,e\kern -0.1em V}}\xspace}
\newcommand{\gevc}{\ensuremath{{\mathrm{\,Ge\kern -0.1em V\!/}c}}\xspace}
\newcommand{\mevc}{\ensuremath{{\mathrm{\,Me\kern -0.1em V\!/}c}}\xspace}
\newcommand{\gevcc}{\ensuremath{{\mathrm{\,Ge\kern -0.1em V\!/}c^2}}\xspace}
\newcommand{\gevgevcccc}{\ensuremath{{\mathrm{\,Ge\kern -0.1em V^2\!/}c^4}}\xspace}
\newcommand{\mevcc}{\ensuremath{{\mathrm{\,Me\kern -0.1em V\!/}c^2}}\xspace}

\def\gsim{{~\raise.15em\hbox{$>$}\kern-.85em
          \lower.35em\hbox{$\sim$}~}\xspace}
\def\lsim{{~\raise.15em\hbox{$<$}\kern-.85em
          \lower.35em\hbox{$\sim$}~}\xspace}

\def\tell1  {TELL1\xspace}
\def\ukl1   {UKL1\xspace}

\newcommand{\eg}{\mbox{\itshape e.g.}\xspace}

%% file: additional-symbols.tex
\def\DDbar   {\ensuremath{\kern -0.1em \stackrel{\kern 0.1em \textsf{\fontsize{5pt}{1em}\selectfont(---)}}{D}\kern -0.3em}\xspace}
\def\AAbar   {\ensuremath{\kern -0.2em \stackrel{\kern 0.2em \textsf{\fontsize{5pt}{1em}\selectfont(---)}}{A}\kern -0.3em}\xspace}
\def\AfAfbar   {\ensuremath{\kern -0.2em \stackrel{\kern 0.2em \textsf{\fontsize{5pt}{1em}\selectfont(---)}}{A}_{\kern -0.3em f}\kern -0.3em}\xspace}
\def\dacp     {\ensuremath{\Delta A_{\CP}}\xspace}
\def\deltam   {\ensuremath{\delta m}\xspace}
\def\besiii {BESIII\xspace}
\def\Mbar    {\kern 0.2em\overline{\kern -0.2em \PM}{}\xspace}
\def\M       {\ensuremath{\PM}\xspace}
\def\Mz      {\ensuremath{\M^0}\xspace}
\def\Mzb     {\ensuremath{\Mbar^0}\xspace}

%% file: cpv_intro.tex
\section{\CP violation in heavy flavour mesons}
\label{s:cpv_intro}
The mass eigenstates of neutral mesons, $|\PM_{1,2}\rangle$, with masses $m_{1,2}$ and widths $\Gamma_{1,2}$, are linear
combinations of the flavour eigenstates, $|\Mz\rangle$ and $|\Mzb\rangle$, as $|\PM_{1,2}\rangle=p|\Mz\rangle\pm{}q|\Mzb\rangle$ with
complex coefficients satisfying $|p|^2+|q|^2=1$.
This allows the definition of the averages $m\equiv(m_1+m_2)/2$ and $\Gamma\equiv(\Gamma_1+\Gamma_2)/2$.
The phase convention of $p$ and $q$ is chosen such that $\CP|\Mz\rangle=-|\Mzb\rangle$.

Following the notation of~\cite{Kagan:2009gb}, the time dependent decay rates of \Mz and \Mzb decays to the final state $f$ can be expressed as
\begin{align}
\label{eqn:charm_start}
\Gamma&(\Mz(t)\to f)=\frac{1}{2}e^{-\tau}\left|A_f\right|^2\nonumber\\*
& \times\Big\{ \left(1+|\lambda_f|^2\right)\cosh(y\tau)+\left(1-|\lambda_f|^2\right)\cos(x\tau) \nonumber\\*
&\quad+2\Re(\lambda_f)\sinh(y\tau)-2\Im(\lambda_f)\sin(x\tau)\Big\},\nonumber\\
\Gamma&(\Mzb(t)\to f)=\frac{1}{2}e^{-\tau}\left|\bar{A}_f\right|^2\nonumber\\*
& \times\Big\{ \left(1+|\lambda^{-1}_f|^2\right)\cosh(y\tau)+\left(1-|\lambda^{-1}_f|^2\right)\cos(x\tau) \nonumber\\*
&\quad+2\Re(\lambda^{-1}_f)\sinh(y\tau)-2\Im(\lambda^{-1}_f)\sin(x\tau)\Big\},
\end{align}
where $\tau\equiv\Gamma t$, \AfAfbar~~are the decay amplitudes and $\lambda_f$ is given by
\begin{equation}
\label{eqn:charm_lambda}
\lambda_f\equiv\frac{q\bar{A}_f}{pA_f}=-\eta_{\CP}\left|\frac{q}{p}\right|\left|\frac{\bar{A}_f}{A_f}\right|e^{i\phi},
\end{equation}
where the right-hand expression is valid for a \CP eigenstate $f$ with eigenvalue $\eta_{\CP}$ and $\phi$ is the \CP violating relative phase between $q/p$ and $\bar{A}_f/A_f$.

In general, \CP symmetry is violated if $\lambda_f$, as defined in Equation~\ref{eqn:charm_lambda}, deviates from $1$.
This can have different origins: the case $|q/p|\neq1$ is called \CP violation in mixing, $|\bar{A}_f/A_f|\neq1$ is \CP violation in
the decay, and a non-zero phase $\phi$ between $q/p$ and $\bar{A}_f/A_f$ causes \CP violation in the interference between mixing and decay.
Mixing is common to all decay modes and hence \CP violation originating in this process is universal which is called indirect \CP violation.
Decay-specific \CP violation is called direct \CP violation.
An excellent discussion on the different types of \CP violation can be found in section 7.2.1 of~\cite{Sozzi:2008zza}.
As opposed to the strange and the beauty system, \CP violation has not yet been discovered in the charm system,
though the \lhcb collaboration has recently found first evidence for \CP violation in two-body \Dz decays~\cite{Aaij:2011in}.

In the charm system one defines the differences $\Delta{}m_{D}\equiv{}m_2-m_1$ and $\Delta\Gamma_{D}\equiv\Gamma_2-\Gamma_1$.
Furthermore, the mixing parameters are defined as $x\equiv\Delta{}m/\Gamma$ and $y\equiv\Delta\Gamma/(2\Gamma)$.
Analogously, in the beauty system one defines the differences $\Delta{}m_{d,s}\equiv{}m_2-m_1$ and $\Delta\Gamma_{d,s}\equiv\Gamma_1-\Gamma_2$, where
the subscripts denote the $\B^0_{d}$ and $\B^0_s$ systems, respectively.

\begin{figure}
\includegraphics[width=0.45\textwidth]{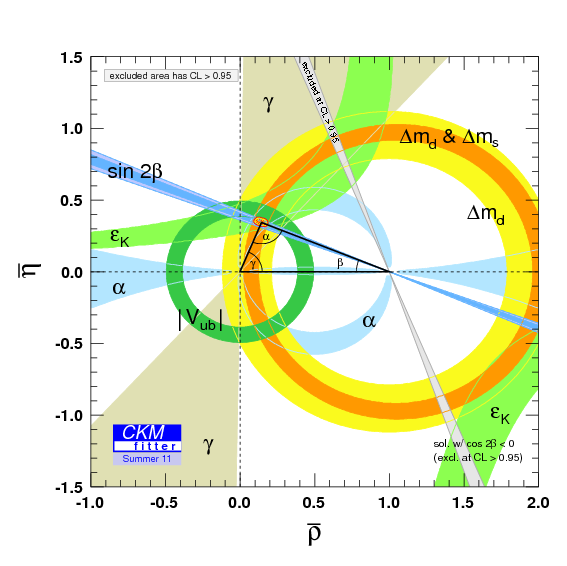}
\caption{The current constraints on the Unitarity Triangle. These meet at the overconstrained apex, and the
shaded ellipse indicates the allowed region for the apex when all measurements are taken together. Reproduced from~\cite{Charles:2004jd}.}
\label{fig:utpict}
\end{figure}

Within the Standard Model (SM), quark mixing is described by the CKM matrix
\begin{multline}
\left(\begin{array}{ccc}V_{ud} & V_{us} & V_{ub} \\ V_{cd} & V_{cs} & V_{cb} \\ V_{td} & V_{ts} & V_{tb}\end{array}\right) = \\
\left(\begin{array}{ccc}1 - \frac{1}{2}\lambda^2 & \lambda & A\lambda^3\left(\rho-i\eta\right) \\ 
-\lambda & 1 - \frac{1}{2}\lambda^2 & A\lambda^2 \\ 
A\lambda^3\left(1-\rho+i\eta\right) & -A\lambda^2 & 1\end{array}\right),
\end{multline}
given on the right in the Wolfenstein parametrization where $\lambda\approx 0.22$ is the sine of the Cabibbo angle.
\CP violation then arises solely from the imaginary term in this
matrix. Since the matrix is unitary, it can be represented by six
triangles in the complex plane, defined by unitarity conditions
such as 
\begin{equation}
V^*_{ub}V_{ud} + V^*_{cb}V_{cd} + V^*_{tb}V_{td} = 0,
\label{eq:uni}
\end{equation}
which is known as the ``Unitarity Triangle''. This particular unitarity condition is chosen because the three terms, corresponding
to the sides of the triangle, are of approximately equal size. The fact that the SM predicts $O(10\%)$ \CP violating effects
in many \B decays, while the predictions for \D decays are generally at least two orders of magnitude smaller, has
led to differing experimental approaches. In the case of \B decays, the focus has been on precise measurements of mixing and \CP violation in order
to overconstrain the sides and angles of the Unitarity Triangle, in particular its apex, as illustrated in Fig.~\ref{fig:utpict}.
In \D decays the focus
has been on searches for \CP violation and a precise understanding of the mixing parameters.

%% file: charm.tex
\section{Charm mixing and \CP violation}
\label{sec:cp_charm}
The studies of charm mesons have gained in momentum with the measurements of first evidence for meson anti-meson mixing in neutral charm mesons in 2007~\cite{Aubert:2007wf,Staric:2007dt}.
Mixing of \Dz mesons is the only mixing process where down-type quarks contribute to the box diagram.
Unlike \PB-meson mixing where the top-quark contribution dominates, the third generation quark is of similar mass to the other down-type quarks.
This leads to a combination of GIM cancellation~\cite{Glashow:1970gm} and CKM suppression~\cite{Cabibbo:1963yz,Kobayashi:1973fv}, which results in a strongly suppressed mixing process~\cite{Bobrowski:2010xg}.

Since experimental evidence has shown that quan\-tum-loop effects are accessible in the charm sector, measurements of \PD mesons provide access to effects from particles beyond the SM, complementary to measurements in the \PB sector.
It was discussed whether the measured size of the mixing parameters could be interpreted as a hint for new physics~\cite{Hou:2006mx,Ciuchini:2007cw,Nir:2007ac,Blanke:2007ee,He:2007iu,Chen:2007yn,Golowich:2007ka}.
New physics effects were also searched for in numerous \CP-violation measurements, which are covered in the remainder of this section, and searches for rare decays as discussed in section~\ref{sec:rare}.

\subsection{Charm mixing}
Mixing of \Dz mesons can be measured in several different modes.
All require identifying the flavour of the \Dz at production as well as at the time of the decay.
Tagging the flavour at production usually exploits the strong decay \decay{\Dstarp}{\Dz\pip}, where the charge of the pion determines the flavour of the \Dz.
Charge conjugate decays are implicitly included here and henceforth.
The small amount of free energy in this decay leads to the difference in the reconstructed invariant mass of the \Dstarp and the \Dz, $\deltam\equiv{}m_{\Dstarp}-m_{\Dz}$, exhibiting a sharply peaking structure over a threshold function as background.
An alternative to using this decay mode is tagging the \Dz flavour by reconstructing a flavour-specific decay of a \PB meson.
This method has not yet been used in a measurement as it did not yet yield competitive quantities of tagged \Dz mesons.
At \lhcb this approach may be of interest due to differences in trigger efficiencies partly compensating for lower production rates.
Another option available, particularly at $\ep\en$ colliders, is the reconstruction of the opposite side charm meson in a flavour specific decay.

Theoretically, the most straight-forward mixing measurement is that of the rate of the forbidden decay \decay{\Dz}{\Kp\mun\neumb} which is only accessible through \Dz-\Dzb mixing.
The ratio of the time-integrated rate of these forbidden decays to their allowed counterparts, \decay{\Dz}{\Km\mup\neum}, determines $R_{\rm m}\equiv(x^2+y^2)/2$.
As this requires very large samples of \Dz mesons no measurement has thus far reached sufficient sensitivity to see evidence for \Dz mixing.
The most sensitive measurement to date has been made by the \belle collaboration to $R_{\rm m}=(1.3\pm2.2\pm2.0)\times10^{-4}$~\cite{Bitenc:2008bk}, where the first uncertainty is of statistical and the second is of systematic nature.
This notation is applied to all results where two uncertainties are quoted.

Related to the semileptonic decay is the suppressed decay \decay{\Dz}{\Kp\pim}, called the wrong-sign (WS) decay.
For this decay, a doubly Cabibbo-suppressed (DCS) amplitude interferes with the decay through a mixing process followed by the Cabibbo-favoured (CF) decay \decay{\Dz}{\Km\pip}.
Following from equation~\ref{eqn:charm_start} the time-dependent decay rate of the WS decay is, in the limit of \CP conservation, proportional to~\cite{Bergmann:2000id}
\begin{equation}
\frac{\Gamma(\Dz(t)\to \Kp\pim)}{e^{-\Gamma t}}\propto \left(R_{\rm D}+\sqrt{R_{\rm D}}y'\Gamma{}t+R^2_{\rm m}(\Gamma{}t)^2\right),
\end{equation}
where the mixing parameters are rotated by the strong phase between the DCS and the CF amplitude, leading to the observable $y'=y\cos\delta_{\PK\Ppi}-x\sin\delta_{\PK\Ppi}$.
The parameter $R_{\rm D}$ is the ratio of the DCS to the CF rate.
Measurements with sufficient sensitivity to unveil evidence for \Dz mixing have been performed by the \babar and \cdf collaborations, leading to $x'^2=(-0.22\pm0.30\pm0.20)\times10^{-3}$ and $y'=(9.7\pm4.4\pm3.1)\times10^{-3}$~\cite{Aubert:2007wf}, and $x'^2=(-0.12\pm0.35)\times10^{-3}$ and $y'=(8.5\pm7.6)\times10^{-3}$~\cite{Aaltonen:2007uc}, respectively.

Similarly, the CF and DCS amplitudes can also lead to higher mass states of the same quark content.
The decay \decay{\Dz}{\Km\pip\piz} is the final state of several such resonances.
Thus, by studying the decay-time dependence of the various resonances a mixing measurement can be obtained.
The \babar collaboration achieved a measurement showing evidence for \Dz mixing with central values of $x''=(26.1^{+5.7}_{-6.8}\pm3.9)\times10^{-3}$ and $y''=(-0.6^{+5.5}_{-6.4}\pm3.4)\times10^{-3}$~\cite{Aubert:2008zh}, where the rotation between the observables and the system of mixing parameters is given by a strong phase as $x''=x\cos\delta_{\Km\pip\piz}+y\sin\delta_{\Km\pip\piz}$ and $y''=y\cos\delta_{\Km\pip\piz}-x\sin\delta_{\Km\pip\piz}$.

The strong phases are not accessible in these measurements but have to come from measurements performed using quantum-correlated \Dz-\Dzb pairs produced at threshold.
These are available from \cleo~\cite{PhysRevD.73.034024,PhysRevD.77.019901,Rosner:2008fq,Asner:2008ft} and can be further improved by \besiii.

By the time of this review no single experiment observation of mixing in \Dz mesons with a significance exceeding $5\sigma$ has been possible.
However, the combination of the numerous measurements by the Heavy Flavor Averaging Group (HFAG) excludes the no-mixing hypothesis by about $10\sigma$~\cite{Asner:2010qj}.
Under the assumption of no \CP violation the world average of the mixing parameters is $x=(6.5^{+1.8}_{-1.9})\times10^{-3}$ and $y=(7.3\pm1.2)\times10^{-3}$.

\subsection{Charm \CP violation}
Indirect \CP violation is often measured in conjunction with mixing parameters.
One example is the measurement of effective inverse lifetimes in decays of \Dz (\Dzb) mesons into final states which are \CP eigenstates, $\hat{\Gamma}$ ($\hat{\bar{\Gamma}}$).
The comparison of these lifetimes to that of a Cabibbo-favoured flavour eigenstate ($\Gamma$) leads to the observable
\begin{align}
\ycp&=\frac{\hat{\Gamma}+\hat{\bar{\Gamma}}}{2\Gamma}-1\nonumber\\
&\approx\eta_{\CP}\left[\left( 1 -\frac{A_m^2}{8}\right)y\cos\phi -\frac{A_m}{2}x\sin\phi\right],
\end{align}
where $A_m$ is the \CP violation in mixing defined alongside the direct \CP violation $A_d$ by $|\lambda_f^{\pm1}|^2\approx(1\pm A_m)(1\pm A_d)$~\cite{Gersabeck:2011xj}.
In the limit of \CP conservation \ycp equals the mixing parameter $y$.

Comparing the \CP eigenstates $\Km\Kp$ and $\pim\pip$ to the Cabibbo-favoured mode $\Km\pip$, the \belle and \babar collaborations have measured $\ycp=(13.1\pm3.2\pm2.5)\times10^{-3}$~\cite{Staric:2007dt} and $\ycp=(11.6\pm2.2\pm1.8)\times10^{-3}$~\cite{PhysRevD.80.071103}, respectively.
The \belle collaboration has also published a measurement using only the decay \decay{\Dz}{\KS\Km\Kp} in which they compare the effective lifetime around the \Pphi resonance with that measured in sidebands of the $\Km\Kp$ invariant mass.
The effective \CP eigenstate content in these regions is determined with two different models.
Their result is $\ycp=(1.1\pm6.1\pm5.2)\times10^{-3}$~\cite{Zupanc:2009sy}.
Provided measurements of sufficient precision, the comparison of \ycp with the mixing parameter $y$ is a test of \CP violation.
However, while one would expect $\ycp<y$ in the presence of \CP violation, the experimental results currently favour $\ycp>y$, i.e.\ no sign of \CP violation is observed.

A second, more sensitive, way of measuring indirect \CP violation is through the comparison of effective lifetimes of \Dz and \Dzb decays to \CP eigenstates.
This leads to the observable
\begin{equation}
\agamma=\frac{\hat{\Gamma}-\hat{\bar{\Gamma}}}{\hat{\Gamma}+\hat{\bar{\Gamma}}}\approx\eta_{\CP}\left[\frac{1}{2}\left( A_m+A_d\right)y\cos\phi -x\sin\phi\right],
\end{equation}
which has contributions from both direct and indirect \CP violation~\cite{Kagan:2009gb,Gersabeck:2011xj}.
Currently there are three measurements of \agamma, which are all compatible with zero.
The \belle, \babar, and \lhcb collaborations have measured $\agamma=(0.1\pm3.0\pm1.5)\times10^{-3}$~\cite{Staric:2007dt}, $\agamma=(2.6\pm3.6\pm0.8)\times10^{-3}$~\cite{PhysRevD.78.011105}, and $\agamma=(-5.9\pm5.9\pm2.1)\times10^{-3}$~\cite{Aaij:2011ad}, respectively.
With the \lhcb result being based only on a small fraction of the data recorded so far, significant improvements in sensitivity may be expected in the near future.
Using current experimental bounds values of \agamma up to $\mathcal{O}(10^{-4})$ are expected~\cite{Kagan:2009gb,Bigi:2011re}.
It has however been shown that enhancements up to about one order of magnitude are possible for example in the presence of a fourth generation of quarks~\cite{Bobrowski:2010xg} or in a Little Higgs Model with T-Parity~\cite{Bigi:2011re}. 
This would bring \agamma close to the current experimental limits.

Eventually, the interpretation of \CP violation results requires precise knowledge of both mixing and \CP violation parameters.
The analysis of the decays \decay{\Dz}{\KS\pim\pip} and \decay{\Dz}{\KS\Km\Kp} offers separate access to the parameters $x$, $y$, $|q/p|$ and $\arg(q/p)$.
This require the decay-time dependence of the phase space structure of these decays, which is possible in two ways: using Dalitz plot models or based on a measurement of the strong phase difference across the Dalitz plot by the \cleo collaboration~\cite{Libby:2010eq}.
One measurement made by the \belle collaboration has determined these parameters based on a Dalitz plot model~\cite{Abe:2007rd}.
Other measurements were performed by the \cleo~\cite{Asner:2005sz} and \babar~\cite{delAmoSanchez:2010xz} collaborations assuming \CP conservation and thus extracting only $x$ and $y$.
With the data samples available and being recorded at \lhcb and those expected at future flavour factories, these measurements will be very important to understand charm mixing and \CP violation.
However, in order to avoid systematic limitations it will be important to reduce model uncertainties or to improve model-independent strong phase difference measurements, which is possible at \besiii.

\begin{figure}
\includegraphics[width=0.45\textwidth]{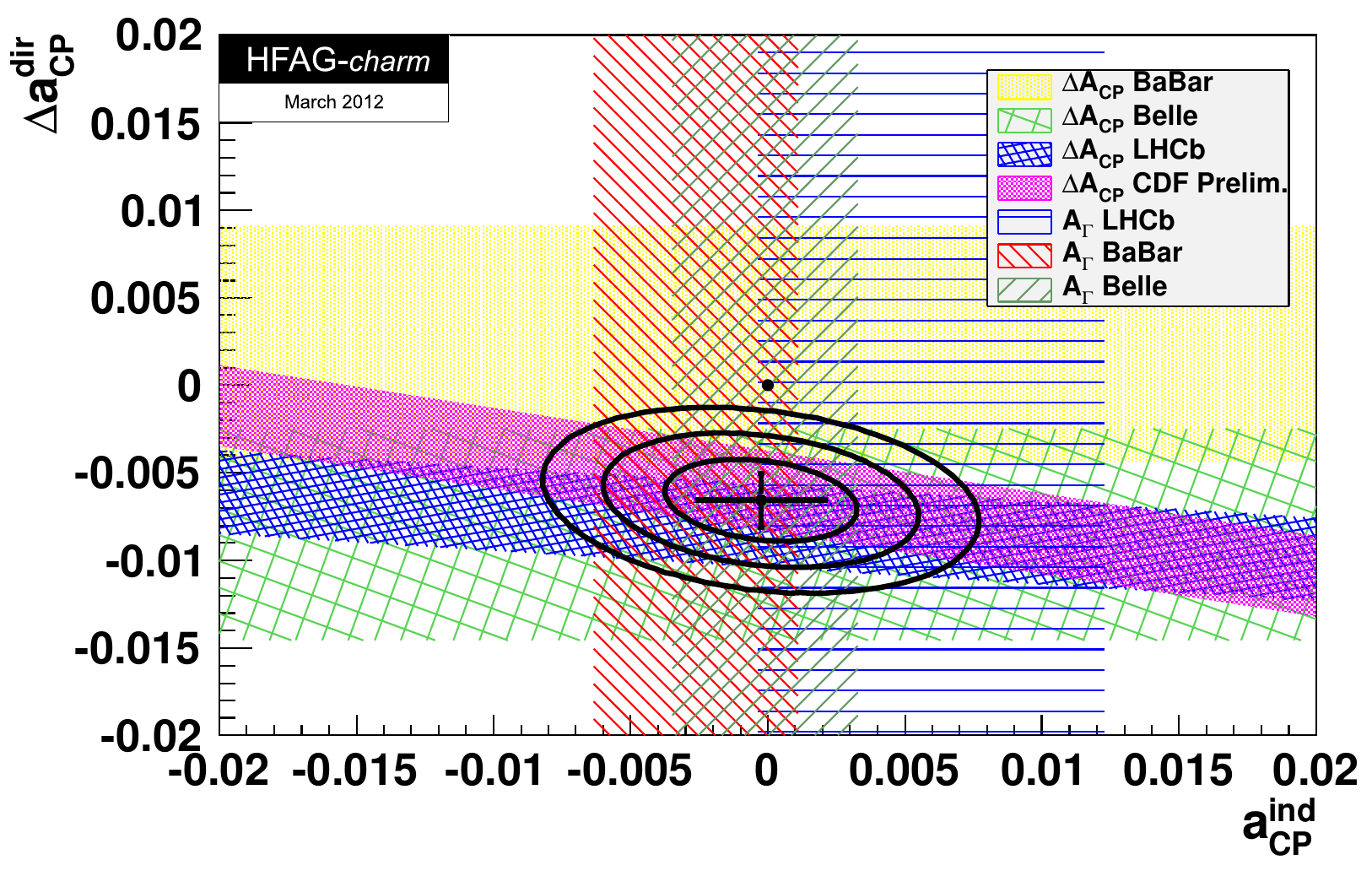}
\caption{HFAG combination of measurements of \dacp and \agamma. Shown are the experimental results as bands indicating their $\pm1\sigma$ uncertainties, the best fit value with one-dimensional uncertainties as a cross, and the $1\sigma$, $2\sigma$, and $3\sigma$ ellipses. The dot marks the point of no \CP violation. Reproduced from~\cite{Asner:2010qj}.}
\label{fig:hfag-charm}
\end{figure}

Direct \CP violation is searched for in decay-time integrated measurements.
However, the decay-time distribution of the data has to be taken into account to estimate the contribution from indirect \CP violation.
Currently, the most striking measurements have been made in decays of \Dz mesons into two charged pions or kaons.
While early measurements of \babar~\cite{Aubert:2007if} and \belle~\cite{Staric:2008rx} had not shown significant deviations from zero, \lhcb recently reported first evidence for \CP violation in the charm sector~\cite{Aaij:2011in}
\begin{align*}
\dacp&\equiv{}A_{\CP}(\Km\Kp)-A_{\CP}(\pim\pip)\nonumber\\
&=(-8.1\pm2.1\pm1.1)\times10^{-3}.
\end{align*}
Meanwhile, \cdf has released a preliminary measurement of $\dacp=(-6.2\pm2.1\pm1.0)\times10^{-3}$~\cite{CDF10784} which shows a hint of a deviation from zero, in support of the \lhcb result.
The observable \dacp exploits the cancellation of systematic uncertainties in the difference of asymmetries.
It gives access to the difference in direct \CP violation of the two decay modes through
\begin{equation}
\dacp=\Delta{}a_{\CP}^{\rm dir}\left(1+\ycp\frac{\overline{\langle{}t\rangle}}{\tau}\right)+\overline{A}_\Gamma\frac{\Delta\langle{}t\rangle}{\tau},
\end{equation}
where $\tau$ is the nominal \Dz lifetime, $\overline{X}\equiv(X(\Km\Kp)+X(\pim\pip))/2$, and $\Delta{}X\equiv{}X(\Km\Kp)-X(\pim\pip)$~\cite{Gersabeck:2011xj}.
With the current precision on \agamma the influence of direct \CP violation on \agamma can be neglected as it is known to be $\le10^{-4}$ and hence $\agamma=-a_{\CP}^{\rm ind}$ is assumed.
Thus, the world average leads to central values of $\Delta{}a_{\CP}^{\rm dir}=(-6.6\pm1.5)\times10^{-3}$ and $a_{\CP}^{\rm ind}=(-0.3\pm2.3)\times10^{-3}$ which has a confidence level of being in agreement with the no \CP violation hypothesis of $6.1\times10^{-5}$~\cite{Asner:2010qj} (see Fig.~\ref{fig:hfag-charm}).

While it was commonly stated in literature that \CP violation effects in these channels were not expected to exceed $10^{-3}$, this statement has been revisited in numerous recent publications.
To date, no clear understanding of whether~\cite{Brod:2011re,Feldmann:2012js,Bhattacharya:2012ah,Franco:2012ck} or not~\cite{Bigi:2011re,Rozanov:2011gj,Cheng:2012wr,Li:2012cf} \CP violation of this level can be accommodated within the SM has emerged.
In parallel with attempts to improve the SM calculations, many estimates of potential effects of PBSM have been made~\cite{Bigi:2011re,Feldmann:2012js,Rozanov:2011gj,Grossman:2006jg,Isidori:2011qw,Wang:2011uu,Hochberg:2011ru,Pirtskhalava:2011va,Chang:2012gn,Giudice:2012qq,Altmannshofer:2012ur,Chen:2012am,Gedalia:2012pi}.
To complement theoretical calculations, measurements in related modes have been and will be performed in order to single out effects from particular amplitudes.

A related way of searching for \CP violation is using decays of charged \PD mesons.
One group of measurements studies decays of \Dp and \Ds mesons into three charged hadrons, namely pions or kaons.
Here, \CP violation can occur in two-body resonances contributing to these decay amplitudes.
Asymmetries in the Dalitz-plot substructure can be measured using an amplitude model or using model-independent statistical analyses~\cite{Bediaga:2009tr,Williams:2011cd}.
The latter allow \CP asymmetries to be discovered while eventually a model-dependent analysis is required to identify its source.
Neither phase-space integrated asymmetry measurements~\cite{Aitala:1996sh,Link:2000aw,Aubert:2005gj,Alexander:2008aa,Rubin:2008aa,Mendez:2009aa}, nor searches for local asymmetries in the Dalitz plot~\cite{Aubert:2005gj,Rubin:2008aa,Staric:2011en,PhysRevD.84.112008,PhysRevLett.108.071801} have shown any evidence for \CP violation.
The largest signal is the recently reported measurement of \CP violation in \decay{\Dp}{\Pphi\pip} of $A_{\CP}^{\Pphi\pip}=(5.1\pm2.8\pm0.5)\times10^{-3}$ by the \belle collaboration~\cite{PhysRevLett.108.071801}, which exploits cancellation of uncertainties through a comparison of asymmetries in the decays of \Dp and \Ds mesons into the final state $\Pphi\pip$.

Decays of \Dp and \Ds into a \KS and either a \Kp or a \pip are closely related to their \Dz counterparts.
Measurements of time-integrated asymmetries in these decays are expected to exhibit a contribution from \CP violation in the kaon system.
As pointed out recently~\cite{Grossman:2011to} this contribution depends on the decay-time acceptance of the \KS.
This can lead to different expected values for different experiments which so far has not been taken into account.
Measurements of asymmetries in the decays \decay{\Dp}{\KS\pip}~\cite{PhysRevD.80.071103,Mendez:2009aa,PhysRevLett.104.181602} and \decay{\Ds}{\KS\pip}~\cite{Mendez:2009aa,PhysRevLett.104.181602} show significant asymmetries.
Future, more precise measurements will reveal whether or not these are in agreement with the expected contribution from the kaon system.

In the light of the recent measurements it is evident that there are four directions to pursue: more precise measurements of \dacp and the individual asymmetries are required to establish the effect; further searches for time-integrated \CP violation need to be carried out in a large range of modes that allow to identify the source of the \CP asymmetry; searches for time-dependent \CP asymmetries, particularly via more precise measure\-ments of \agamma; and finally a more precise determination of the mixing parameters is required.
Complementary to this are searches for rare charm decays, studies of the top quark~\cite{Wang:2011uu,Hochberg:2011ru}, measurements of nuclear electric dipole moments~\cite{Giudice:2012qq}, and many other flavour observables which are beyond the scope of this review.

%% file: BCPV.tex
\section{Beauty mixing and \CP violation}
\label{sec:cpv_beauty}
The existence of \Bd and \Bs meson mixing is well established, and the mass difference between the light
and heavy eigenstates has been measured to high precision in both systems. In addition, evidence exists
for \CP violation in \Bd, \Bu, and \Bs decays. The interpretation of the experimental data focuses on the compatibility
of the various measurements with each other, and their compatibility with the SM description of \CP violation as arising from
a single weak phase in the CKM matrix.
Two tensions stand out at present:
the discrepancy between the large mixing-induced \CP asymmetry measured in semileptonic \Bz and \Bs decays~\cite{Abazov:2011yk}
and the small \CP violating phase in \Bs mixing~\cite{LHCb-CONF-2012-002} on the one hand, and the 
discrepancy between $\sin\left(2\beta\right)$ and \Vub measured from the branching ratio of $\Bp\to\taup\nu$~\cite{Asner:2010qj}
on the other hand. 
\subsection{\Bs mixing}
The mixing of \Bs mesons is described by the width difference between the light and heavy mass eigenstates, $\Delta\Gamma_s$,
the mass difference $\Delta m_s$, 
and a single \CP violating phase $\phi_s$. Within the SM the width difference is substantial,
$\Delta\Gamma_s = \Gamma_{\textrm{L}} - \Gamma_{\textrm{H}} = 0.087\pm0.021$~ps$^{-1}$~\cite{Lenz:2011ti},
while the \CP violating phase, as determined from indirect fits to experimental data,
is small $\phi_s = -0.036\pm0.002$~rad~\cite{Lenz:2011ti,Lenz:2006hd,Charles:2011va}.
Both can deviate substantially from these predictions in other models. 

The first observation of \Bs mixing was made by \cdf~\cite{Abulencia:2006ze}, while
the most precise measurement of the mass difference $\Delta m_s$ comes from the recent
\lhcb measurement~\cite{Aaij:2011qx}.
The most precise measurements
of both the width difference and phase come from the measurement of the time-dependent \CP asymmetry in
\BsToJPsiPhi~\cite{LHCb-CONF-2012-002,CDF-CONF-10778,PhysRevD.85.032006}
\begin{align*}
\phi_s &=& -0.001 \pm 0.101 \pm 0.027 \;\textrm{rad}\textrm{~\cite{LHCb-CONF-2012-002}},\\
\Delta\Gamma_s &=& 0.116 \pm 0.018 \pm 0.006 \;\textrm{ps}^{-1}\textrm{~\cite{LHCb-CONF-2012-002}},\\
\phi_s &\in& [\frac{\pi}{2},-1.51]\cup [-0.06,0.30]\cup [1.26,\frac{\pi}{2}]\;\textrm{rad}\textrm{~\cite{CDF-CONF-10778}},\\
\phi_s &=& -0.55^{+0.38}_{-0.36}\;\textrm{rad}\textrm{~\cite{PhysRevD.85.032006}}.
\end{align*}
All these measurements are in good agreement with the SM, and it is notable that a non-zero $\Delta\Gamma_s$ has
been directly measured for the first time at $5\sigma$. In addition, the sign of $\Delta\Gamma_s$ has been unambiguously
determined to be positive through the study of S-wave and P-wave
contributions to the $\Bs\to\jpsi K^+K^-$ decay amplitude~\cite{LHCb-PAPER-2011-028}.

The measurement of $\phi_s$ from \BsToJPsiPhi is complicated by the
vector-vector final state, which necessitates a time-dependent angular analysis, whereas it was proposed~\cite{Stone:2008ak} to study the
vector-pseudoscalar decay $\Bs\to\jpsi f_0(980)$ in which no such analysis is required. This measurement has recently been
performed by the \lhcb collaboration, which, combined with the \lhcb \BsToJPsiPhi measurement leads to
\begin{equation*}
\phi_s = -0.002 \pm 0.083 \pm 0.027 \;\textrm{rad}\;\textrm{\cite{LHCb-CONF-2012-002}}\; ,
\end{equation*}
in good agreement with the SM prediction.

As noted in~\cite{Fleischer:2010ib},
the interplay of $\Delta\Gamma_s$ and $\phi_s$ leads to predictions for the effective lifetimes of \Bs mesons decaying
into \CP eigenstates. In the specific case of $\Bs\to K^+ K^-$, the lifetime has already been measured~\cite{Aaij:2011kn,LHCb-CONF-2012-001} to be 
$1.468 \pm 0.046 \pm 0.006 \textrm{ps}$. Using the latest measurement of
$\Gamma_s$ and $\Delta\Gamma_s$ by \lhcb~\cite{LHCb-CONF-2012-002}, as well as the \Bs lifetime $\tau_{\Bs} = 1.472 \pm 0.025$~ps~\cite{Asner:2010qj},
the SM prediction from~\cite{Fleischer:2010ib} can be updated to $\tau_{K^+ K^-} = 1.40 \pm 0.02$.
Moreover, recent first observations of $\Bs\to\Dz\Dz$ and $\Bs\to\Dp\Dm$~\cite{LHCb-CONF-2012-009} by \lhcb
indicate that it will be possible in the near future to measure effective lifetimes in many different \Bs decays to \CP eignestates,
and further constrain $(\phi_s,\Delta\Gamma_s)$ in this manner.

The decay $\Bs\to K^+ K^-$ is not only a decay to a \CP eigenstate, but is one example of a $b\to s$ penguin transition in
the decays of \Bs mesons. 
One of the experimentally most interesting modes of this type is $\Bs\to \phi\phi$ where, because of a cancellation of \CP violating effects from decay and mixing,
the SM predicts an upper limit
of $0.02$ for \CP violation~\cite{Raidal:2002ph}. Although the time-dependent analysis
is yet to be performed, time-integrated analyses based on measuring triple products have been performed, and have found
no significant asymmetries~\cite{Aaltonen:2011rs,LHCb-PAPER-2012-004}, in agreement with SM predictions.

Another interesting~\cite{Ciuchini:2007hx} decay is $\Bs\to \Kstarz \Kstarz$, which has recently
been observed for the first time by \lhcb~\cite{LHCb-PAPER-2011-012}. Because of the $V-A$ structure of the weak interaction,
the \CP-even longitudinal polarization component was expected to be dominant~\cite{Beneke:2006hg,Cheng:2009mu,Ali:2007ff} in both this decay and $\Bs\to \phi\phi$.
However, both B-factory measurements in $b\to s$ penguin modes~\cite{Abe:2004mq,Chen:2005zv,Aubert:2006uk,Aubert:2006fs,Aubert:2008zza,delAmoSanchez:2010mz},
as well as the recent \lhcb measurements of
$\Bs\to \phi\phi$ and $\Bs\to \Kstarz \Kstarz$, find roughly equal longitudinal and \CP-odd transverse polarization
components. Proposed explanations have included large penguin annihilation contributions~\cite{Kagan:2004uw} or final state interactions~\cite{Datta:2007qb}.
The time dependent \CP violation measurements in both these modes should
become experimentally accessible in the near future, further
constraining PBSM.
\subsection{\Bd mixing}
The mixing of \Bd mesons can be described within the same formalism as that of \Bs mesons, but now it is the width difference
$\Delta\Gamma_d$ which is small in the SM while the mixing phase $\phi_d$ is large. The most
precise measurements of $\Delta m_d$ were made by \babar~\cite{Aubert:2005kf} and \belle~\cite{Abe:2004mz},
leading to the current world average $\Delta m_d = 0.505\pm0.004$~\cite{Nakamura:2010zzi}.
The mixing phase can also be expressed
as the angle $\beta$ of the Unitarity Triangle, whose most precise measurement comes from the study of time-dependent
\CP violation in the ``golden mode'' $\Bd\to\jpsi\KS$ and related decays 
\begin{align*}
\sin(2\beta) &=& 0.687 \pm 0.028 \pm 0.012\textrm{~\cite{Asner:2010qj,Aubert:2009aw}},\\
\sin(2\beta) &=& 0.667 \pm 0.023 \pm 0.012\textrm{~\cite{Adachi:2012et}},
\end{align*}
The measurement of this angle can be related to the CKM matrix element \Vub through the unitarity relation in equation~\ref{eq:uni},
and can be compared to the value of $\sin(2\beta)$ as determined from a fit to the other 
parameters of the Unitarity Triangle~\cite{Charles:2004jd,Ciuchini:2000de} of $0.830^{+0.013}_{-0.033}$
and $0.80\pm 0.05$ from the CKMFitter and UTFit collaborations respectively.
This tension is driven by the branching fraction of the decay $\Bp\to\taup\nu$
\begin{align*}
B(\B\to\tau\nu) &=& (1.80^{+.57}_{-.54} \pm 0.26)\times 10^{-4}\textrm{~\cite{delAmoSanchez:2010ab}},\\
B(\B\to\tau\nu) &=& (1.54^{+.38}_{-.37} {}^{+.29}_{-.31})\times 10^{-4}\textrm{~\cite{Hara:2010dk}},
\end{align*}
which can be transformed into a measurement of \Vub and hence a constraint on the apex of the Unitarity Triangle.

Resolving this tension will require a precise understanding of the size of doubly Cabibbo-suppressed
penguin topologies in $\Bd\to\jpsi\KS$~\cite{Faller:2008zc}.
In this respect it is interesting to note the observation of the U-spin partner
decay $\Bs\to\jpsi\KS$ at \lhcb~\cite{LHCb-CONF-2011-048}, which has been proposed~\cite{Fleischer:1999nz,DeBruyn:2010hh} as one
way of measuring these effects. 

An important additional null-test of the SM comes from the measurement of $\Delta\Gamma_d$. As noted in~\cite{Gershon:2010wx},
the fact that the SM prediction for $\Delta\Gamma_d/\Gamma_d$ is so small, $40.9^{+8.9}_{-9.9}\times 10^{-4}$~\cite{Lenz:2006hd},
while plausible scenarios of PBMS exist in which this value is enhanced~\cite{Lenz:2012az}, means that any non-zero
measurement with current experimental sensitivity would be a clear sign of new physics effects. Indeed, such effects are
needed to explain the anomalous dimuon asymmetry observed by \dzero, as discussed in the following section. 
Both \babar and \belle have measured $\Delta\Gamma_d$~\cite{Aubert:2003hd,Aubert:2004xga,Higuchi:2012kx} through
fits to the time dependent decay rates in $\Bd\to\D^{(*)-} (\pi,\rho,a_1)^{+}$ and $\Bd\to c\bar{c} K^0_{\textrm{S,L}}$ modes.
The average is dominated by the recent \belle result of 
$\Delta\Gamma_d/\Gamma_d = [-1.7\pm 1.8 \pm 1.1]\times 10^{-2}$. As the uncertainty on this
measurement is still an order of magnitude larger than the SM prediction, it remains to be seen if the
systematic uncertainties can be kept under control in the era of the next generation flavour factories.
\subsection{Semileptonic asymmetries}
The mixing induced semileptonic asymmetry $A_{\textrm{sl}}$ is predicted to be $O(10^{-4})$ in the SM within
both the \Bz $(a^d_{\textrm{sl}})$ and \Bs $(a^s_{\textrm{sl}})$ meson systems~\cite{Lenz:2011ti}. The most precise experimental measurement to
date was made by the \dzero Collaboration~\cite{Abazov:2011yk}, which found a percent-level \CP asymmetry
\begin{align*}
A_{\textrm{sl}} &\approx& 0.6\times a^d_{\textrm{sl}} + 0.4\times a^s_{\textrm{sl}}, \\
A_{\textrm{sl}} &=& (-0.787 \pm 0.172 \pm 0.093) \%\;.
\end{align*}
Because \dzero cannot distinguish between dimuon pairs coming from \Bz and \Bs decays, it measures a combination of the
two semileptonic asymmetries. In the same paper, the collaboration attempts to separate effects caused by \Bs oscillations
from those caused by \Bz oscillations by indirectly studying the lifetime of the decaying \B meson, and concludes that the
asymmetry is largest at short lifetimes. The authors take this as a hint that the asymmetry is dominated by \Bs decays
because the \Bs meson oscillates much more quickly than the \Bz.

When interpreting this result, it is important to keep in mind that the background levels are also highest at short
lifetimes; for this reason, it is critical that $(a^d_{\textrm{sl}})$ and $(a^s_{\textrm{sl}})$ are measured separately in
a low background environment where the decaying \B meson can be unambiguously tagged as a \Bz or \Bs.
Nevertheless, taking the \dzero result at face value, it is not trivial to reconcile it with the measurements of \Bs
and \Bd mixing mentioned earlier. An easy way of seeing this is to consider why, if the dimuon asymmetry is driven by \Bs
mixing, the mixing phase in \BsToJPsiPhi is so close to the SM value
while the direct and indirect measurements of $\sin\left(2\beta\right)$ are in tension.
One proposed explanation requires~\cite{Lenz:2012az} contributions from PBSM to both $M^{d,s}_{\textrm{12}}$ and
$\Gamma^{d,s}_{\textrm{12}}$
\subsection{$B \to h \gamma$ decays}
CP asymmetry measurements of $b\to s \gamma$
transitions are sensitive to PBSM, for instance through
measurements of the photon polarisation which probes models involving right-handed
currents~\cite{Gershon:2006mt,DescotesGenon:2011yn,Mahmoudi:2006wq,Altmannshofer:2011gn}. 
\CP asymmetries in $b \to s \gamma$ transitions have
been measured by \babar~\cite{Aubert:2008gy,Aubert:2008be,Aubert:2009ak},
\belle~\cite{Nishida:2003yw,Ushiroda:2006fi}, and \lhcb~\cite{LHCb-CONF-2012-004} and the results are consistent with
SM expectations and statistically limited. 
In this context, it has been recently noted~\cite{Benzke:2010tq} that the difference
in \CP asymmetries between the inclusive processes $X_s^{+}\gamma$ and $X_s^{0}\gamma$ offers a cleaner probe of
PBSM than either measurement taken on its own.

Thanks to the large value of $\Delta \Gamma_s$, the $B_s$ system
is particularly promising for measuring the photon
polarisation by studying time dependent \CP violation in the decay $B_s \to \phi \gamma$~\cite{Muheim:2008vu}. This mode
was first observed at \belle~\cite{Wicht:2007ni}, while 
the \lhcb collaboration has recently measured~\cite{LHCb:2012ab} the ratio of the
branching ratios 
$\frac{{\cal B} (B^0 \to K^{*} \gamma)}{B_s \to \phi \gamma)} = 1.12
\pm 0.08^{+0.11}_{-0.09}$.
\subsection{The CKM angle $\gamma$}
A precise determination of the angle $\gamma$ of the CKM unitarity triangle is important in order to further overconstrain
the position of the triangle's apex, in particular with respect to the previously discussed measurements of
$\sin\left(2\beta\right)$ and $V_{ub}$. In this respect $\gamma$ can be measured either from tree-level or loop-mediated
processes, and a comparison of the two kinds of measurements provides another opportunity for PBSM to manifest itself.
In either case, $\gamma$ is experimentally determined from a measurement of \CP violation in
those \B meson decays where diagrams involving \Vub and \Vcb result in the same final state.

The determination of $\gamma$ from tree-level decays is one of the most sensitive tests of the SM precisely because
the associated theoretical uncertainties are confined to electroweak corrections associated with box-diagram decays,
and are at the level of $\delta\gamma/\gamma \approx 10^{-6}$~\cite{Zupan:2011mn}. Experimentally the challenge is that
the sensitivity to $\gamma$ comes from the interference of \Vub and \Vcb diagrams, which means that the final state must
be carefully chosen in order to make the amplitudes of similar size and hence maximize the interference. Unfortunately those
modes which have the highest interference also have the biggest associated experimental difficulties, whether it be low
overall branching ratios, difficult to reconstruct final state particles, or the requirement for a time-dependent analysis.
This means that the ultimate precision on $\gamma$ can only be achieved by combining several different measurements.

The current sensitivity on $\gamma$ is dominated by measurements of \CP violation and partial widths in
$\Bp\to\Dz\Kp$ decays, in which the \Dz then decays to either
a \CP-eigenstate~\cite{Gronau1991483,Gronau1991172},
a doubly-Cabibbo suppressed decay mode~\cite{Atwood:1996ci,Atwood:2000ck}, or a multibody decay whose Dalitz distribution
gives rise to interference effects~\cite{Giri:2003bs}. These are known as the GLW, ADS, and GGSZ methods respectively after their inventors.

In the first two cases the charge-averaged partial
width ratios of the $\Dz\Kp$ and $\Dz\pip$ decays are measured,
\begin{equation}
R^f_{K/\pi} = \frac{\Gamma(\B\to [f]_D K)}{\Gamma(\B\to [f]_D \pi)},
\end{equation} 
where $f$ represents the \CP-eigenstate $\pi\pi$ and $KK$ decays and the Cabibbo-favoured $K\pi$ decay mode; the \CP 
asymmetries 
\begin{equation}
A^f_h = \frac{\Gamma(\Bp\to [f]_D h^+) - \Gamma(\Bm\to [f]_D h^-) }{\Gamma(\Bp\to [f]_D h^+) + \Gamma(\Bm\to [f]_D h^-)},
\end{equation} 
where $h$ is a pion or a kaon; and the charge-separated partial width ratios of the Cabibbo-favoured and doubly Cabibbo-suppressed
$\Bp\to\Dz\Kp$ decay modes 
\begin{equation}
R^\pm_{h} = \frac{\Gamma(\B^\pm\to [\pi^\pm K^\mp]_D h^\pm)}{\Gamma(\B^\pm\to [K^\pm \pi^\mp]_D h^\pm)}.
\end{equation} 
As these are the most experimentally accessible modes for measuring $\gamma$, they have been studied at \babar~\cite{delAmoSanchez:2010dz,delAmoSanchez:2010ji},
\belle~\cite{Abe:2006hc,Belle:2011ac},\cdf~\cite{Aaltonen:2009hz,Aaltonen:2011uu}, and recently at \lhcb~\cite{Aaij:2012kz}.
In particular, \lhcb has observed the doubly Cabibbo-suppressed decay
$\Bpm \to [\pi^\pm K^\mp]_D K^\pm$ with $10\sigma$ significance, and has made a $5.8\sigma$ observation of \CP-violation in $\Bp\to\Dz \Kp$
decays. It is worth highlighting the cleanliness of the \lhcb signals, as seen in Fig.~\ref{fig:gammab2dk},
as well as the intriguing hint
of \CP-violation in the $\Bpm \to [\pi^\pm K^\mp]_D \pi^\pm$ which can be seen in the same picture. 
\begin{figure}
\includegraphics[width=0.45\textwidth]{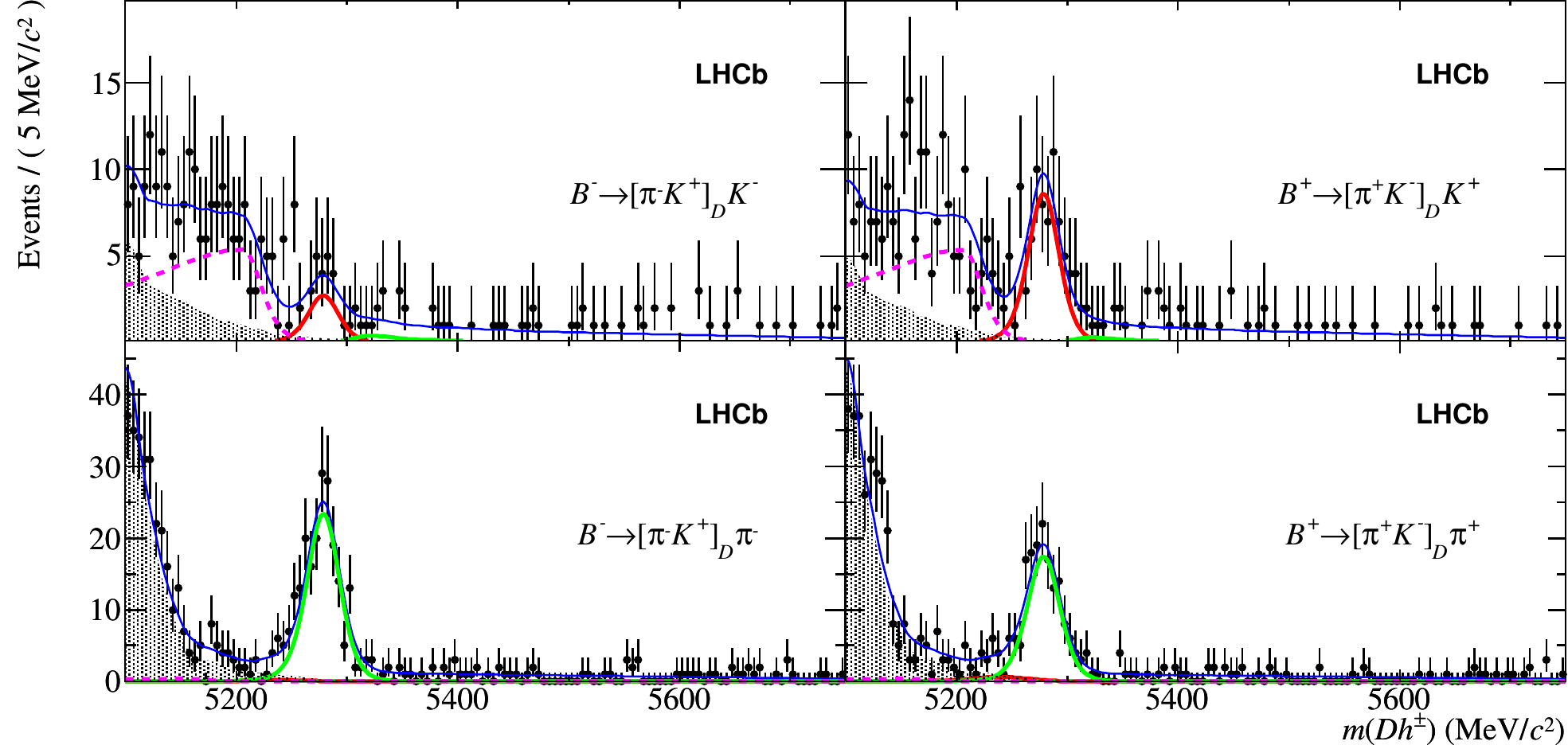}
\caption{Invariant mass distribution of selected $\Bpm \to [\pi^\pm K^\mp]_D h^\pm$ candidates. The left plots are \Bm, the right
plots are \Bp. Top are $h=K$ and bottom are $h=\pi$. The red curve is the signal, the shaded area, green,
and magenta curves are backgrounds. Reproduced from~\cite{Aaij:2012kz}.}
\label{fig:gammab2dk}
\end{figure}

In the third case, what is measured are the different Dalitz plot distributions of $\Dz\to \KS hh$ in
$\Bp\to \Dz \Kp$ and $\Bm\to\Dz \Km$ decays, and measurements have been made
with~\cite{delAmoSanchez:2010rq,Poluektov:2010wz} or without~\cite{Belle:2011ab} assuming
an amplitude model for the \Dz decay. The advantage of this method is that it only suffers from a two-fold
ambiguity in the measured value of $\gamma$, as opposed to the eightfold ambiguity in \eg the GLW method.

The average value of $\gamma$ from these
decay modes, as computed by the CKMFittter collaboration,
is shown in Fig.~\ref{fig:gammab2dk}, from which it is apparent that while direct measurements of $\gamma$ agree well
with its indirect determination from other Unitarity Triangle parameters, they are not yet strongly constraining
the apex of the triangle. A historical tension exists between the frequentist (CKMFitter) and Bayesian (UTFit) averages of $\gamma$, driven
by the different treatment of the nuissance parameters which parametarize the size of the interference in each decay mode.
The most up-to-date averages from the two collaborations are
\begin{align*}
\gamma &=& (66 \pm 12)^\circ &\;\textrm{CKMFitter}&\; \textrm{\cite{Charles:2004jd}}\; , \\
\gamma &=& (76 \pm 9)^\circ &\;\textrm{UTFit}&\; \textrm{\cite{Ciuchini:2000de}}\; ,
\end{align*}
where the CKMFitter average includes the most recent ADS/GLW results from \lhcb and the UTFit average does not.
The larger uncertainty in the CKMFitter average comes from the treatment of the nuissance parameters, while there
is an interesting discrepancy developing in the central values which is not understood at present.

\begin{figure}
\includegraphics[width=0.45\textwidth]{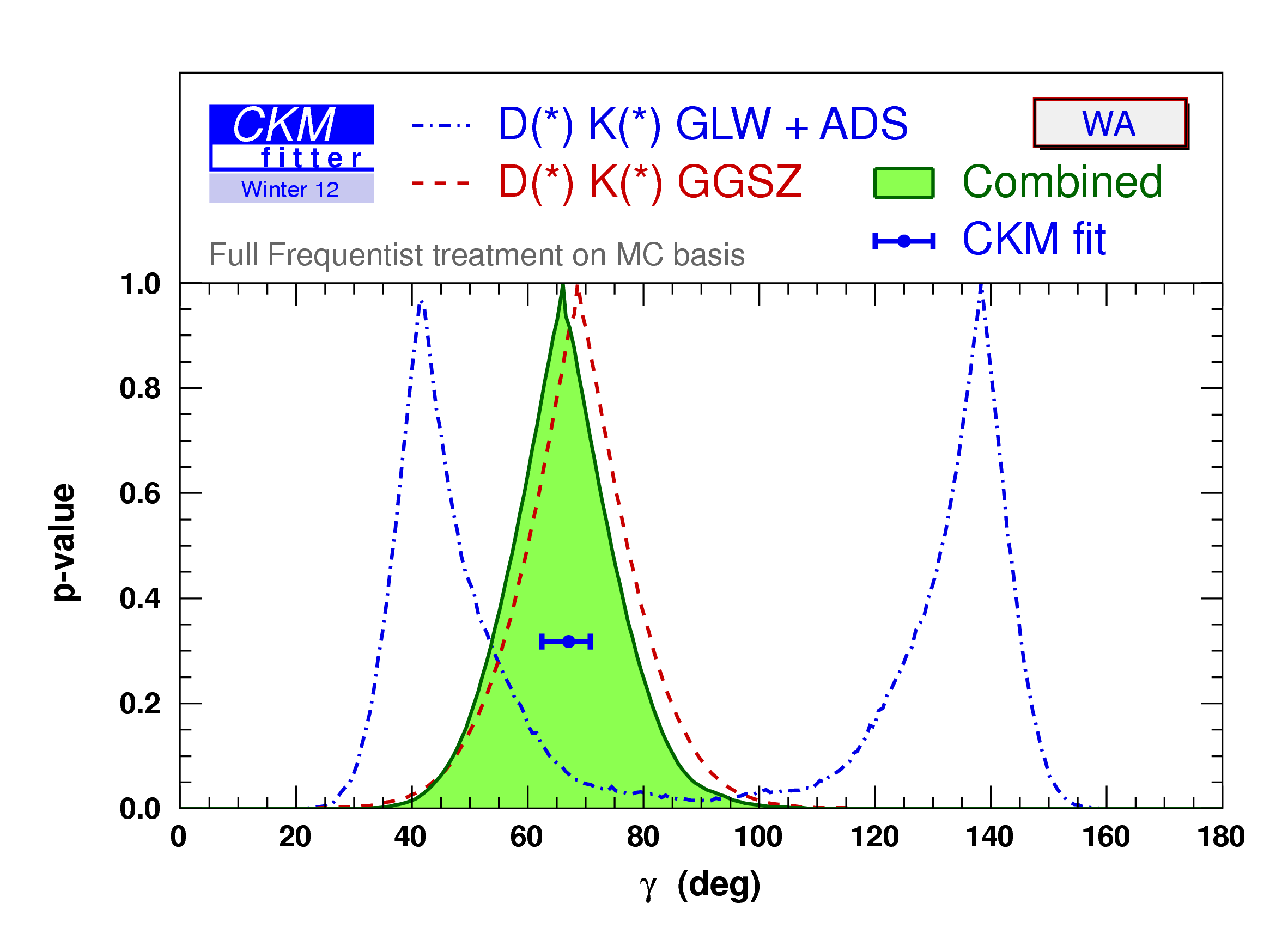}
\caption{Averaged constraints on $\gamma$ from direct measurements. Reproduced from~\cite{Charles:2004jd}.}
\label{fig:gammab2dk}
\end{figure}

Many other tree-level determinations of $\gamma$ are possible, for example from $\Bd\to\Dz h^+h^-$
decays~\cite{Gronau1991483,Gronau1991172,Atwood:1996ci,Atwood:2000ck,Gershon:2008pe,Gershon:2009qc}
whether in a quasi-two-body approach, selecting the $h^+ h^-$ mass to lie at a particular resonance,
or through an amplitude analysis. An important milestone on this road to $\gamma$ is
the first observation of the decay mode $\Bs\to\Dz\Kstarz$ at \lhcb~\cite{Aaij:2011tz}.
It is also possible to make an unambiguous measurement of $\gamma$ through the study of \CP violation in the
interference of \Bs mixing and the decay $\Bs\to\Dspm K^\mp$~\cite{FleischerDsK}, whose branching ratio has recently been precisely
measured~\cite{LHCb-CONF-2011-057}. Within measurements
of $\gamma$ from loop-mediated processes, the study of two body $\B_{s,d}\to h^+ h^-$ decays stands out.
The U-spin partner decays $\Bs\to K^+ K^-$ and $\Bd\to \pi^+ \pi^-$ are able to extract $\gamma$ unambiguously
in a combined analysis~\cite{Fleischer:1999pa,Fleischer:2007hj}, and recently the time-dependent \CP asymmetry in $\Bs\to K^+ K^-$ has been measured~\cite{LHCb-CONF-2012-007} for the first time
\begin{align*}
A^{\textrm{dir}}_{KK} &=& 0.02 \pm 0.18 \pm 0.04\;,\\
A^{\textrm{mix}}_{KK} &=& 0.17 \pm 0.18 \pm 0.05\;,
\end{align*}
to add to the existing~\cite{Aubert:2008sb,Ishino:2006if} measurements in $\Bd\to \pi^+ \pi^-$. 

%% file: RareDecays.tex
\section{Rare Decays}
\label{sec:rare}
Rare decays which proceed via Flavour Changing Neutral Currents
(FCNC) are induced by one-loop diagrams in the SM and are excellent
probes for PBSM.  
New particles can enter in competing loop-order diagrams, resulting in
large deviations from SM predictions. 
In general, an effective hamiltonian formalism is used to describe the amplitudes of
FCNC processes, according to the formula:
\begin{equation}
H_{eff} = \frac{G_F}{\sqrt{2}}\sum_i V_{CKM}^i C_i(\mu) Q_i\;,
\end{equation}   
where $V^i_{CKM}$ are the relevant factors of the CKM matrix; $Q_i$ are local operators; $C_i$ are the corresponding couplings
(Wilson coefficients);  and $\mu$ is the QCD renormalization scale.  
The correlation of different channels, where common Wilson
coefficients contribute, is a powerful tool for searching and understanding the structure of
PBSM. \\
\indent This approach is complementary to direct searches for
PBSM. Moreover indirect searches 
often allow to set more stringent constraints than direct ones. 
For instance, strong lower bounds on the mass of the charged
Higgs in Two-Higgs-Doublets-Models of type II have been obtained from the analysis of
$\overline{B} \rightarrow X \gamma$ decays, where
the SM
prediction ~\cite{Misiak:2006zs} is found in agreement with inclusive measurements performed by the
experiments \babar~\cite{Aubert:2006gg,Aubert:2005cba,Aubert:2007my}, \belle~\cite{Abe:2001hk,Limosani:2009qg} and
\cleo~\cite{Chen:2001fja} (other bounds from $B\to X_s \gamma$ are
discussed in~\cite{Buras:2011zb} and the references therein). \\
\indent As a result of the many measurements performed by the B-factories and more
recently by the \cdf experiment,
our knowledge of suppressed processes has considerably improved
in the last decade. Consequently, constraints
on PBSM have become much stronger. \\
\indent While inclusive measurements are
challenging at hadron colliders, studies of exclusive decays are competitive with $e^+e^-$ machines.
Moreover, hadron colliders have the advantage that all B-hadron species are produced. 
With the start-up of the \lhcb experiment a new round in the
precision measurements of rare decays has begun. 

\subsection{$B_{s,d} \rightarrow \mu^{+} \mu^{-}$ decays}
\begin{figure}[!h]
\centering
\includegraphics[scale=0.35]{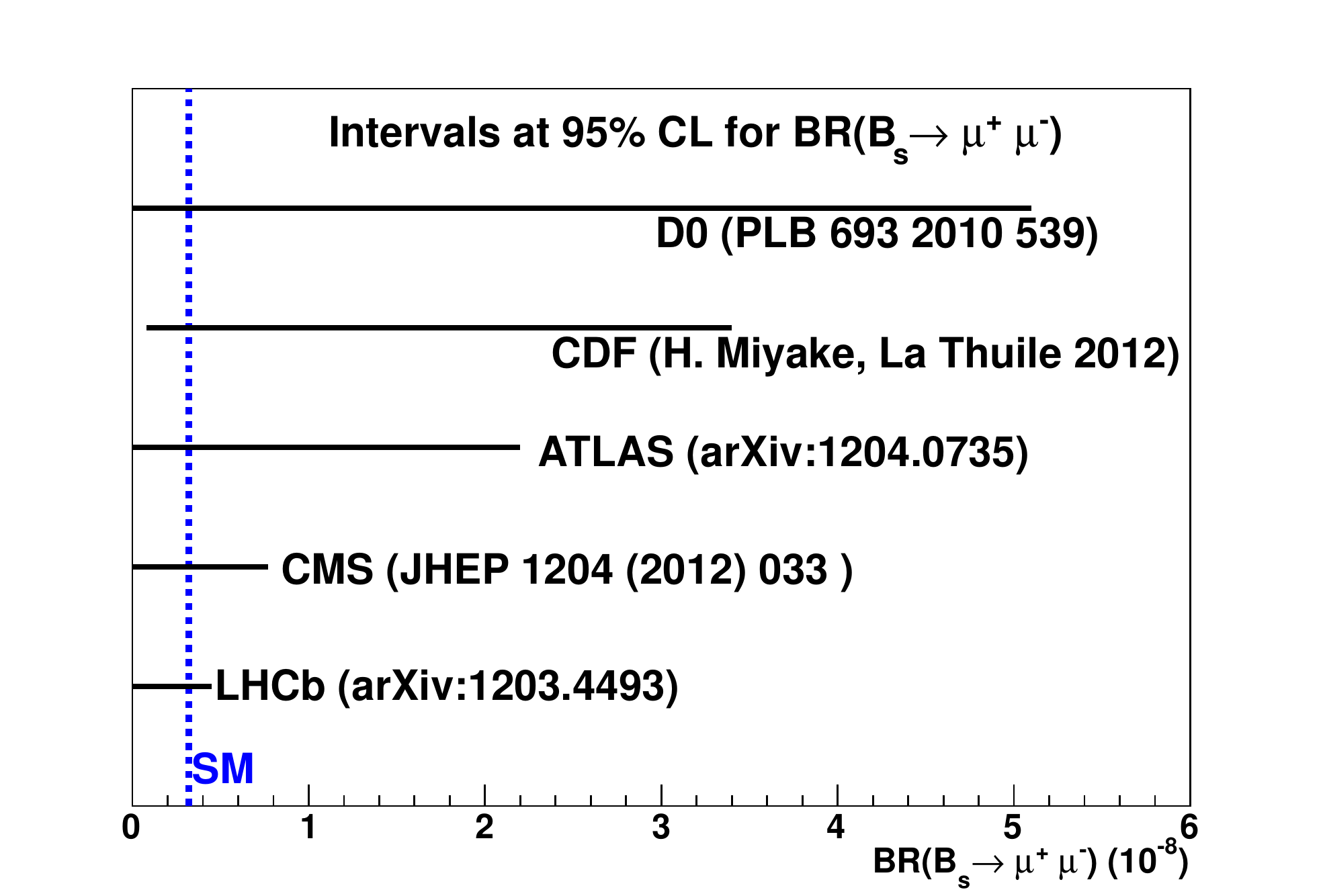}
\caption{Present limits on ${\cal B}(B_s \to \mu^+ \mu^-)$ at $95\%$
  CL set by the experiments D0~\cite{Abazov:2010fs}
  \cdf~\cite{CDF-Bsmumu}, \atlas~\cite{ATLAS-CONF-2012-010}
  \cms~\cite{Chatrchyan:2012rg} and \lhcb~\cite{Aaij:2012ac}. The
  SM prediction is indicated by the blue-dashed line. \label{fig:BsmumuInt}}
\end{figure}

Purely leptonic decays of B-mesons are a key ingredient in the search
for PBSM, since the prediction of their branching fractions is largely free
from hadronic uncertainties. The two decays $B_{s,d} \to \mu^+ \mu^-$
have a clear experimental signature and are easier to reconstruct and
identify than the other leptonic decays of
B-mesons. 
Their branching fractions are predicted to be 
$ {\cal B} (B_s\rightarrow \mu^{+} \mu^{-}) = (3.2 \pm 0.2)\times 10^{-9}$ and 
${\cal B}(B_d\rightarrow \mu^{+} \mu^{-})= (1.0 \pm 0.1)\times
10^{-10}$ in the SM~\cite{Buras:2010mh,Buras:2010wr}. 
Contributions from PBSM, especially in models with an extended Higgs
sector, can enhance these branching fractions. For instance, in the
Minimal Supersymmetric extension of the SM the branching fraction of the
decay $B_s\rightarrow \mu^{+} \mu^{-}$ is proportional to the sixth
power of $\tan \beta$ (the ratio of the vacuum expectation values of
the neutral components of the Higgs fields $H_u$ and
$H_d$)~\cite{Hurth:2008jc}. This fact makes this observable particularly
sensitive to supersymmetric models with large $\tan \beta$. 
More generally measurements of this branching fractions probe the
Wilson coefficients $C_s$ and $C_p$, which are negligibly small in the
SM. Present measurements of ${\cal B}(B_s\rightarrow
\mu^{+} \mu^{-})$ are shown in Fig.~\ref{fig:BsmumuInt}. 

\begin{figure}[!h]
\centering
\includegraphics[scale=0.30]{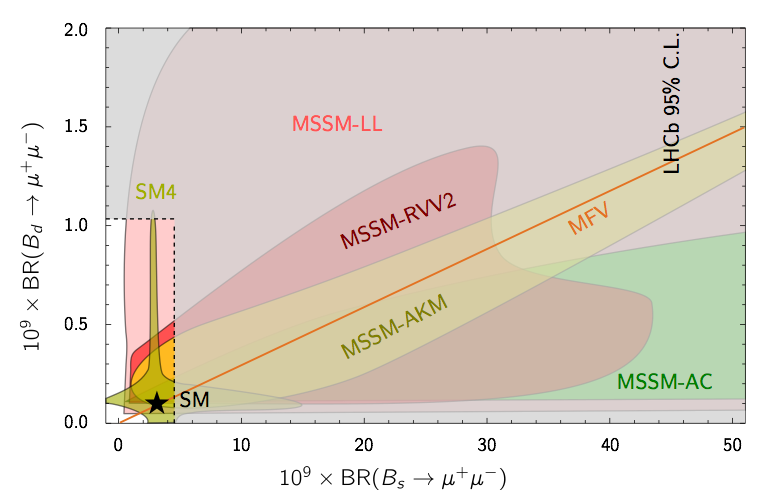}
\caption{Correlation for the branching fractions of the decays $B_s\to
  \mu^+ \mu^-$ and $B_d\to \mu^+ \mu^-$ for several models of PBSM. Details
  on the models can be found in~\cite{Straub:2010ih}. The recent upper limits by \lhcb are
  shown by the shaded region. Reproduced from ~\cite{Straubb-Moriond}. \label{fig:StraubBsMuMu}}
\end{figure}

Presently, the most stringent upper limits on  ${\cal B}(B_{s,d}\rightarrow
\mu^{+} \mu^{-})$  are set by the \lhcb experiment~\cite{Aaij:2012ac}. 
This analysis profits from the good momentum resolution and the good particle identification
performances of \lhcb to reject the different sources of background.
The branching fraction of the signal was extracted by using the three normalization channels: $B^+ \rightarrow J/\psi K^+$, 
$B^0\rightarrow K^+ \pi^-$ and $B_s\rightarrow J/\psi \phi$. For the
first two of these channels, the ratio of the hadronization 
fractions $\frac{f_s}{f_d}$ is needed\footnote{Isospin symmetry, i.e.
  $f_u =f_d$, has been assumed.}. This variable was measured at \lhcb by combining
measurements with semi-leptonic and hadronic
decays~\cite{Fleischer:2010ay}: $f_s/f_d = 0.267^{+0.021}_{-0.020}$~\cite{Aaij:2011jp,Aaij:2011hi}. 
The uncertainty on this parameter is, in the long run,  a limiting
systematic uncertainty for discriminating between SM and BSM
contributions in the $B_s\to
\mu^{+} \mu^{-}$ decay, as well as for the measurement of the
\textit{golden ratio} $\frac{{\cal B} (B_s \to \mu^+ \mu^-)}{{\cal B}
  (B_d \to \mu^+ \mu^-)}$ ~\cite{Buras:2010pi}. The correlation
between the branching fractions of the decays $B_{s,d} \to \mu^+ \mu^-$ is shown
in Fig.~\ref{fig:StraubBsMuMu} for several beyond SM scenarios. 
\indent The upper limits set by \lhcb for the $B_{s,d} \to \mu^+ \mu^-$
decays are: 
${\cal B}(B_s\rightarrow \mu^+ \mu^-)<4.5 \times 10^{-9}$ and ${\cal B}(B_d\rightarrow
\mu^+ \mu^-)<1.05 \times 10^{-9}$ at $95\%$ CL and are illustrated in
Fig.~\ref{fig:StraubBsMuMu} by the shaded region. 
These measurements are in agreement with SM expectations and
give additional constraints for PBSM with respect to those provided by
$b\to s \gamma$ and other $b\to s
l^+l^-$ transitions.

\subsection{$B\to h \mu^+ \mu^-$ decays}

\begin{figure}[!h]
\centering
\includegraphics[scale=0.35]{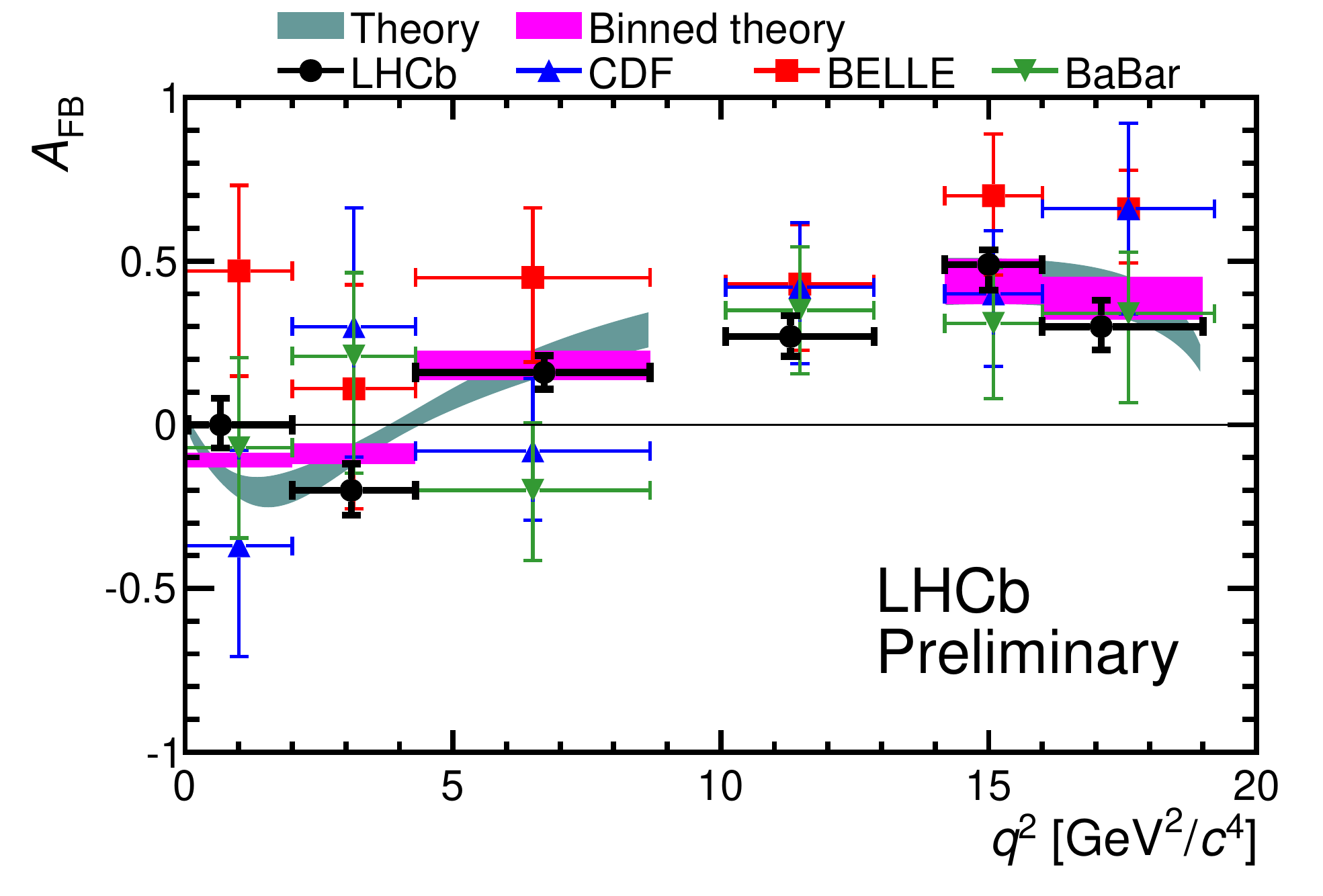}
\includegraphics[scale=0.35]{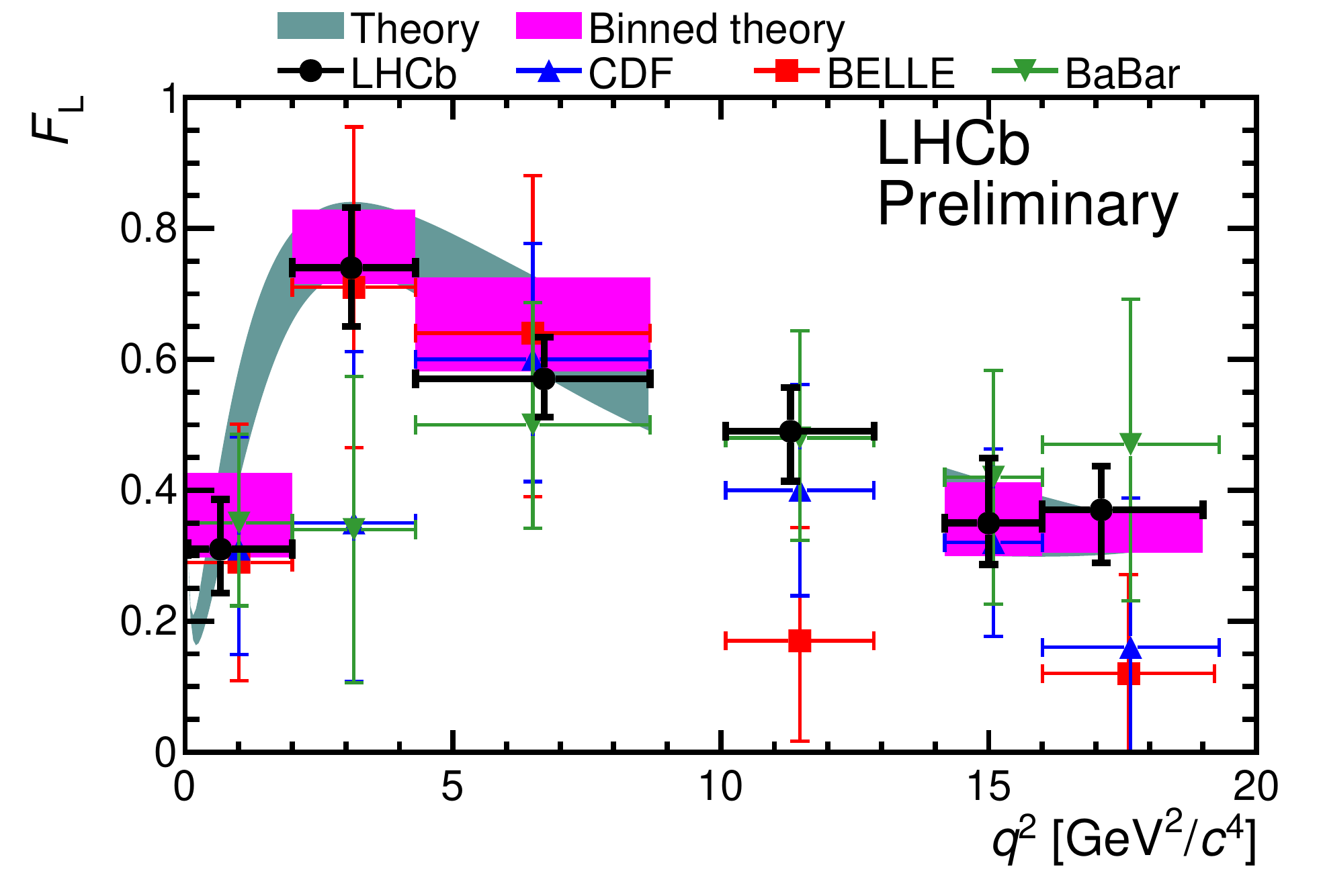}\\
\includegraphics[scale=0.35]{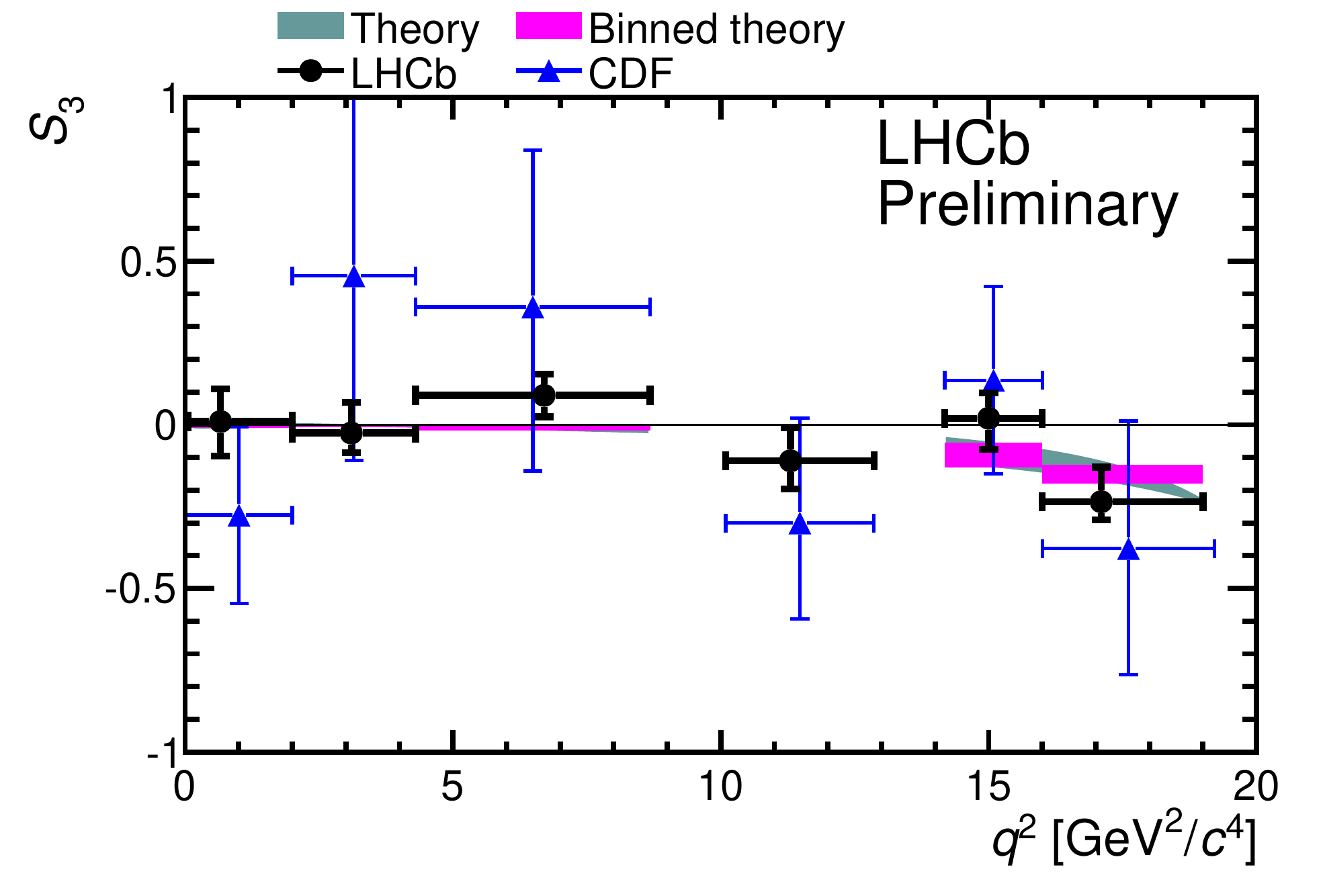}
\includegraphics[scale=0.35]{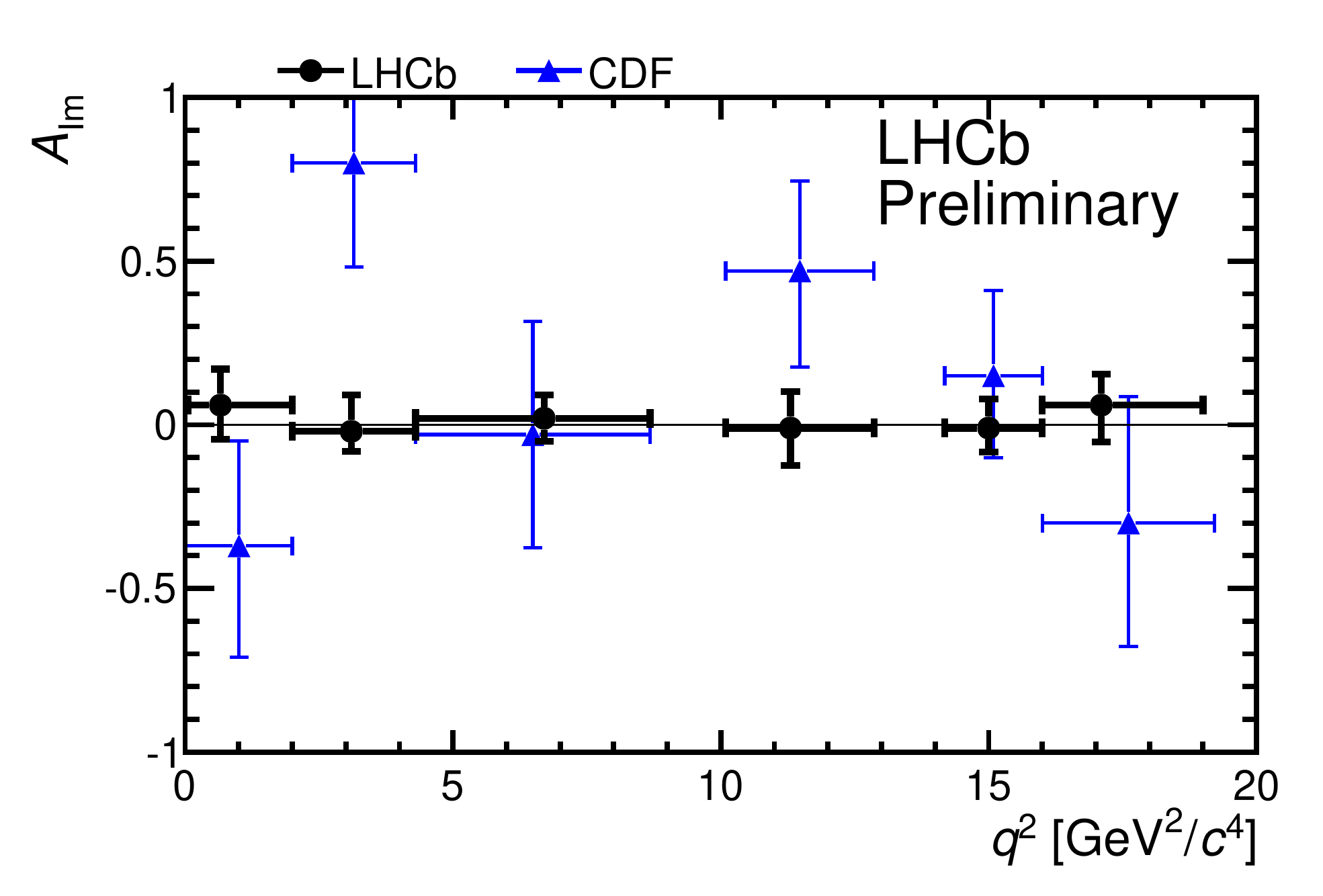}
\caption{The \AFB, $F_L$, $S_3$ and $A_{\textrm{Im}}$ measured by the
  experiments \babar~\cite{BaBarLakeLouise}, \belle~\cite{Wei:2009zv}, \cdf~\cite{Aaltonen:2011ja} and
  \lhcb~\cite{LHCb-CONF-2012-008}. The comparison with the SM prediction, taken
  from~\cite{Bobeth:2011gi} is also shown. Reproduced from ~\cite{LHCb-CONF-2012-008}. \label{fig:Res:Kstmm}}
\end{figure}

In the decay $B_d\rightarrow K^* \mu^+ \mu^-$ several angular
observables can be built which are sensitive to PBSM, and for which form factor uncertainties are 
theoretically under control, (see for example ~\cite{Ali:1991is,Altmannshofer:2008dz} and references therein). 
These observables include the forward-backward asymmetry of the dimuon system, \AFB, the fraction 
of $K^{*}$ longitudinal polarization, $F_L$, the transverse
asymmetry, $S_{3}$ ~\cite{Altmannshofer:2008dz} (often referred to as
$\frac{1}{2}(1-F_L)A_{T}^2$ in the literature~\cite{Kruger:2005ep}), and the T-odd CP
asymmetry $A_{\textrm{Im}}$~\cite{Bobeth:2008ij}. 
They can be extracted by performing an angular analysis as a function of the
dimuon invariant mass squared, $q^2$, with
respect to the following angles: the angle $\theta_l$ between
the $\mu^+$ ($\mu^-$) and the $B^0$ ($\overline{B}^0$) in the dimuon
rest frame; the angle $\theta_K$ between the kaon and $B^0$
in the $K^*$ rest frame; and the  angle $\phi$ between the planes of
the dimuon system and the plane of the $K^*$. A formal definition of
these angles can be found in ~\cite{Egede:1048970}. 
Present measurements of the observables \AFB, $F_L$, $S_3$ and $A_{\textrm{Im}}$  are shown in
Fig.~\ref{fig:Res:Kstmm}. These measurements provide information about the
Wilson coefficients $C_7$, $C_9$ and $C_{10}$ and on their right-handed counterparts. 

The \lhcb experiment  has recently made the world's best measurements on
these angular observables~\cite{LHCb-CONF-2012-008}. 
The physics parameters were extracted by fitting the partial
decay rate as a function of the three angles for different bins in $q^2$.
In order to reduce the number of parameters
in the fit, due to the small size of the data sample, the angle $\phi$ was folded by taking $\phi \to \phi +
\pi$ when $\phi <0$. This transformation cancels out the terms containing $cos \phi$  and $sin \phi$ in
the differential decay rate. 
This strategy is different from that followed by other experiments,
where only projections of the angular distributions were used. 
\begin{figure}[!h]
\centering
\includegraphics[scale=0.35]{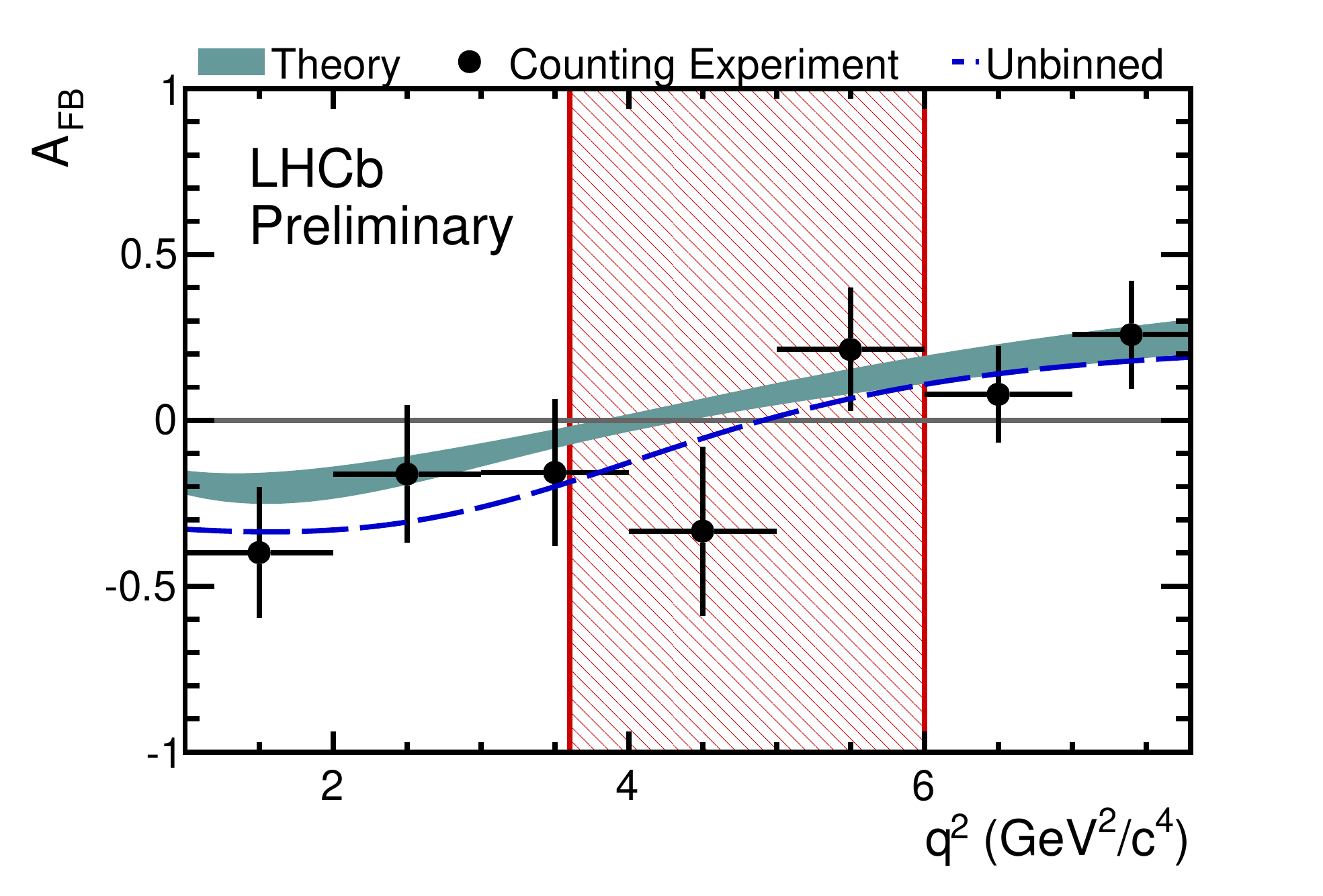}
\caption{The \AFB as a function of $q^2$ extracted from an unbinned
  counting experiment. The shaded region correspond to the $68\%$ CL
  of the zero-crossing point. Comparison with the SM
  prediction~\cite{Bobeth:2011gi} is shown. Reproduced from~\cite{LHCb-CONF-2012-008}.\label{fig:Res:ZCP}}
\end{figure}

\begin{figure}[!h]
\centering
\includegraphics[scale=0.4]{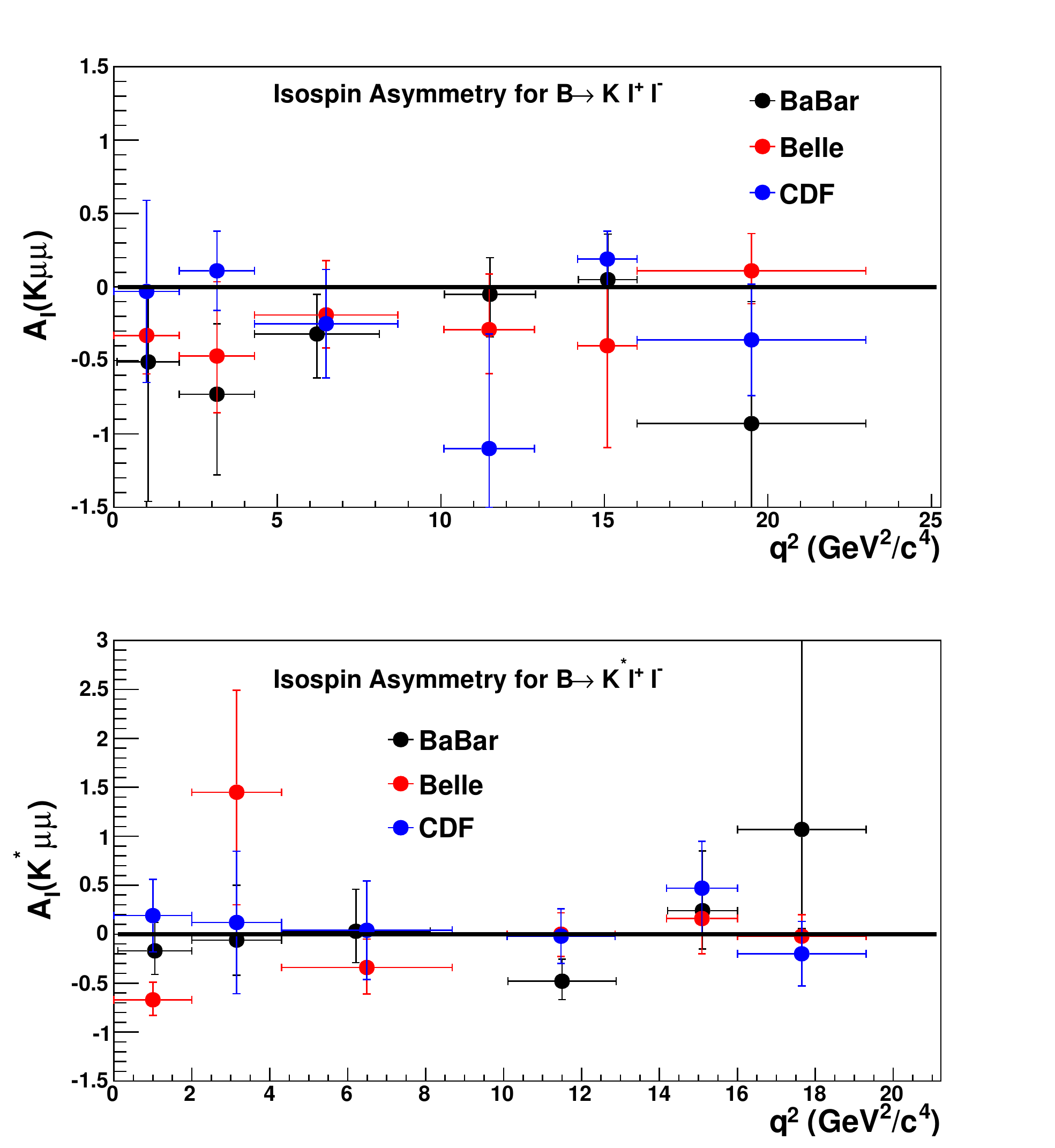}
\caption{Isospin asymmetry for the decays $B\to K^{( * )} l^{+} l^{-}$,
  measured by the experiments \babar~\cite{BaBarLakeLouise},
  \belle~\cite{Wei:2009zv} (with electrons and muons) and
  \cdf~\cite{Aaltonen:2011ja} (with muons).\label{fig:Isospin}}
\end{figure}

\begin{figure}[!h]
\centering
\includegraphics[scale=0.30]{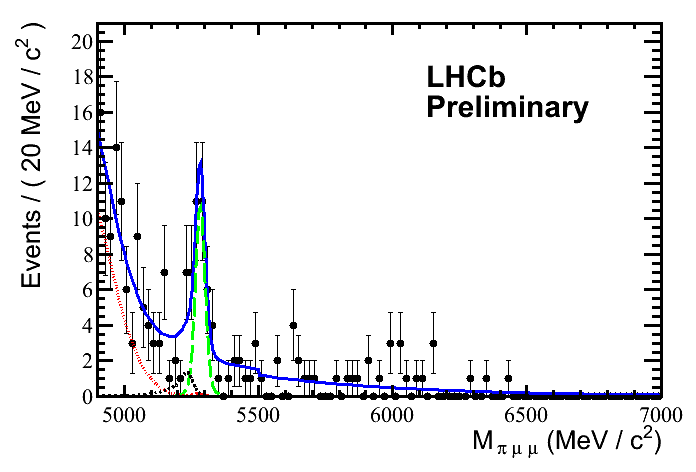}
\caption{The invariant mass distribution of $B^+ \to \pi^+ \mu^+
  \mu^-$ candidates. Reproduced from~\cite{LHCb-CONF-2012-006}. \label{fig:Res:BPimumu}}
\end{figure}

The so called zero-crossing point, where \AFB changes sign, is largely free from form factor
uncertainties and sensitive to 
PBSM~\cite{Ali:1991is}.  The SM predicts this
point to be in the range  4.0-4.3 GeV$^{2}$/c$^{4}$~\cite{Bobeth:2011nj,Beneke:2004dp,Ali:2006ew}. The zero-crossing
point of \AFB was measured
for the first time by \lhcb to be $q_0^2 = 4.9^{+1.1}_{-1.3}$GeV$^{2}$/c$^4$~\cite{LHCb-CONF-2012-008}. 
This observable was extracted in an unbinned counting
  experiment with respect to $q^2$, integrating the angular
distributions with respect to the three angles~\cite{Jansen:1156131}. The result is shown in Fig.~\ref{fig:Res:ZCP}. \\
\indent Other exclusive $b \to s ll$ processes have been measured by the
B-factories, CDF and \lhcb. 
The measurements of the differential branching fractions of the
decays $\Lambda_b \to \Lambda \mu^+ \mu^-$~\cite{Aaltonen:2011qs}, $B^+\rightarrow
K^+\mu^+\mu^-$~\cite{Aaltonen:2011qs,Wei:2009zv,BaBarLakeLouise}, $B_s\rightarrow \phi \mu^+ \mu^-$~\cite{Aaltonen:2011qs}, $B^0\to
K_S\mu^+ \mu^-$~\cite{Aaltonen:2011qs,Wei:2009zv,BaBarLakeLouise} and $B^+\to
K^{*+}\mu^+ \mu^-$~\cite{Aaltonen:2011qs,Wei:2009zv,BaBarLakeLouise}  and the 
$A_{FB}$ for the decays $B^+ \to K^+ \mu^+
\mu^-$~\cite{Aaltonen:2011ja,Wei:2009zv,BaBarLakeLouise} and $B^+\to K^{*+} \mu^+
\mu^-$~\cite{Aaltonen:2011ja,Wei:2009zv,BaBarLakeLouise} were found 
to be in agreement with SM predictions. \\
\indent Another observable which is potentially sensitive to PBSM is the isospin asymmetry, $A_I$, defined as:
\begin{equation}
A_I = \frac{{\cal B}(B^0\to K^{(*)0} l^+
  l^-)-\frac{\tau_0}{\tau_+}{\cal B}(B^{\pm}\to K^{(*)\pm} l^+
  l^-)}{{\cal B}(B^0\to K^{(*)0} l^+ l^-)+\frac{\tau_0}{\tau_+}{\cal
    B}(B^{\pm}\to K^{(*)\pm} l^+ l^-)},
\end{equation} 
where $l=(e, \mu)$. This observable was measured by the experiments 
\belle~\cite{Wei:2009zv}, \babar~\cite{BaBarLakeLouise} and \cdf~\cite{Aaltonen:2011ja}. The results are shown in
Fig.~\ref{fig:Isospin}. 
The SM predicts a small asymmetry in all $q^2$ bins. Present results
seem to hint at a non-zero $A_I$ in some $q^2$ bins. Measurements with
larger data samples and good control of the theoretical uncertainties are important;
a measurement of $A_{I}$ from \lhcb can be expected in the near future. \\
\indent Recently, the \lhcb collaboration reported the first observation of a $b \to d ll$
transition, by measuring the branching fraction  ${\cal B}(B^+ \to
\pi^{+} \mu^+ \mu^-) = (2.4 \pm 0.6 \pm 0.2)\times 10^{-8} $~\cite{LHCb-CONF-2012-006}. 
The invariant mass distribution of $B^+ \to \pi^+ \mu^+ \mu^-$
candidates is shown in Fig.~\ref{fig:Res:BPimumu}.
This process is further suppressed by the factor $|V_{td}/V_{ts}|$, with
respect to the $b \to s ll$ transitions.
The measured branching fraction is in good agreement with the SM expectation. 
   
\subsection{Search for  Lepton Flavour Violating and very rare decays}
The search for Lepton Flavour Violating (LFV) decays is a crucial way
to test the SM flavour structure. 
These searches have been performed by several experiments. A complete discussion of LFV
searches goes beyond the scope of this review. \\
\indent Decays of the type
$B^+\to h^- l^+l^+$, where $h^-$ is a meson, can be considered the
analogues of neutrinoless double $\beta$ decays and
can be used to search for heavy Majorana
neutrinos~\cite{Atre:2009rg,Zhang:2010um,Boyarsky:2006fg,Ilakovac:1999md}. These searches
have been performed by the \lhcb ~\cite{Aaij:2012zr,Aaij:2011ex},
\babar~\cite{BABAR:2012aa}, \belle~\cite{Seon:2011ni} and \cleo
~\cite{Edwards:2002kq} experiments. Upper limits for these decays are
summarised in Table~\ref{tab:exotic:UL}. 
Heavy Majorana neutrinos can also be searched for by using the corresponding
charm decays $D^+\to h^- l^+l^+$. Constraints on these decays are expected
to improve substantially with measurements from \lhcb. \\
\indent LFV decays of charged leptons are allowed in several extensions of the
SM, for instance supersymmetric models~\cite{Dedes:2002rh,Ciuchini:2007ha,Arganda:2005ji}, left-right
symmetric models~\cite{Akeroyd:2006bb} and models with heavy neutrinos~\cite{Atre:2009rg,Zhang:2010um,Boyarsky:2006fg,Ilakovac:1999md}. 
Stringent upper limits on the decay $\mu^- \to e^- \gamma$ have been set by the MEG
experiment~\cite{Adam:2011ch}, while the most stringent upper limits on $\tau^{-} \to
l^- l^+ l^-$ were set by the \belle
experiment~\cite{Hayasaka:2010np}. \\
\indent In addition, searches for exotic very rare decays have been carried out
at \lhcb. Upper limits for the decays $B_{d,s}\to
\mu^+\mu^-\mu^+\mu^-$ and $D^0 \to \mu^+ \mu^-$ were recently
set~\cite{LHCb-CONF-2012-010,LHCb-CONF-2012-005} and are listed in
Table~\ref{tab:exotic:UL}. \\
\indent For the moment no hint of the existence of any of such processes has been observed,
and all searches are statistically limited at present.

\begin{table}[!h]
\centering
\begin{tabular}{|c|c|c|}
\hline
Channel & Upper Limit (CL) & Reference \\
\hline
\hline
${\cal B} (B^+ \to  K^- \mu^+ \mu^+)$ & $5.4 \times 10^{-8}$ ($95\%$) & \lhcb~\cite{Aaij:2011ex} \\
${\cal B} (B^+ \to  \pi^- \mu^+ \mu^+)$ & $1.3 \times 10^{-8}$ ($95\%$) & \lhcb~\cite{Aaij:2012zr} \\
${\cal B} (B^+ \to  \pi^- e^+ e^+)$ & $2.3 \times 10^{-8}$ ($90\%$) & \babar~\cite{BABAR:2012aa} \\
${\cal B} (B^+ \to  K^- e^+ e^+)$ & $3.0 \times 10^{-8}$ ($90\%$) &
\babar~\cite{BABAR:2012aa} \\
${\cal B} (B^+ \to  D^- \mu^+ \mu^+)$ & $6.9\times 10^{-7}$ ($95\%$)
& \lhcb~\cite{Aaij:2012zr} \\
${\cal B} (B^+ \to  D^{*-} \mu^+ \mu^+)$ & $2.8\times 10^{-6}$ ($95\%$)
& \lhcb~\cite{Aaij:2012zr} \\
${\cal B} (B^+ \to  D^- e^+ e^+)$ & $2.6\times 10^{-6}$ ($90\%$)
& \belle~\cite{Seon:2011ni} \\
${\cal B} (B^+ \to  D^- \mu^+ e^+)$ & $1.8\times 10^{-6}$ ($90\%$)
& \belle~\cite{Seon:2011ni} \\
${\cal B} (B^+ \to  D_s^- \mu^+ \mu^+)$ & $5.8\times 10^{-7}$ ($95\%$) & \lhcb~\cite{Aaij:2012zr} \\
${\cal B} (B^+ \to  D^0 \pi^- \mu^+ \mu^+)$ & $1.5 \times 10^{-6}$ ($95\%$) & \lhcb~\cite{Aaij:2012zr} \\
${\cal B}(D^0 \to \mu^+ \mu^+)$ & $1.3 \times 10^{-8}$ ($95\%$) & \lhcb~\cite{LHCb-CONF-2012-005} \\
${\cal B} (B_s \to \mu^+\mu^-\mu^+\mu^-)$ & $1.3 \times 10^{-8}$ ($95\%$) & \lhcb~\cite{LHCb-CONF-2012-010} \\
${\cal B} (B^0 \to \mu^+\mu^-\mu^+\mu^-)$ & $5.4 \times 10^{-9}$ ($95\%$) & \lhcb~\cite{LHCb-CONF-2012-010} \\
${\cal B} (\tau^- \to \mu^-\mu^+\mu^-)$ & $2.1 \times 10^{-8}$ ($90\%$) & \belle~\cite{Hayasaka:2010np} \\
${\cal B} (\tau^- \to e^- e^+ e^-)$ & $2.7 \times 10^{-8}$ ($90\%$) & \belle~\cite{Hayasaka:2010np} \\
${\cal B} (\tau^- \to e^- \mu^+\mu^-)$ & $2.7 \times 10^{-8}$ ($90\%$) & \belle~\cite{Hayasaka:2010np} \\
${\cal B} (\tau^- \to e^+\mu^- \mu^-)$ & $1.7 \times 10^{-8}$ ($90\%$) & \belle~\cite{Hayasaka:2010np} \\
${\cal B} (\tau^- \to \mu^+ e^- e^-)$ & $1.5 \times 10^{-8}$ ($90\%$) & \belle~\cite{Hayasaka:2010np} \\
${\cal B} (\tau^- \to \mu^- e^+ e^-)$ & $1.8 \times 10^{-8}$ ($90\%$) & \belle~\cite{Hayasaka:2010np} \\
${\cal B} (\mu^- \to e^- \gamma)$ & $2.4 \times 10^{-12}$ ($90\%$) & MEG~\cite{Adam:2011ch} \\
\hline
\end{tabular}
\caption{Upper Limit for several very rare or forbidden decays. \label{tab:exotic:UL}}
\end{table}